\documentclass{aa}
\pdfoutput=1
\usepackage{graphicx}
\usepackage{url}
\usepackage{multirow}
\usepackage{natbib}
\bibpunct{(}{)}{;}{a}{}{,}
\usepackage{float}

\usepackage{color}

\DeclareFontFamily{U}{euc}{}
\DeclareFontShape{U}{euc}{m}{n}{<-6>eurm5<6-8>eurm7<8->eurm10}{}%
\DeclareSymbolFont{AMSc}{U}{euc}{m}{n} 
\DeclareMathSymbol{\umu}{\mathord}{AMSc}{"16}

\begin{document}

\title{SPHERE/SHINE reveals\\ concentric rings in the debris disk of HIP 73145  }
\author{M.~Feldt\inst{1}
\and
J. Olofsson\inst{2,1,3}
\and
A. Boccaletti\inst{4}
\and
A.L. Maire\inst{1}
\and
J. Milli\inst{5}
\and
A. Vigan\inst{6}
\and
M. Langlois\inst{6,7}
\and
Th. Henning\inst{1}
\and
A. Moor\inst{8}
\and
M. Bonnefoy\inst{9}
\and
Z. Wahhaj\inst{6,5}
\and
S. Desidera\inst{10}
\and
R. Gratton\inst{10}
\and
\'A. K\'osp\'al\inst{1,8}
\and
P. Abraham\inst{8}
\and
F. Menard\inst{9}
\and
G. Chauvin\inst{9}
\and
A.M. Lagrange\inst{9}
\and
D. Mesa\inst{10}
\and
G. Salter\inst{6}
\and
E. Buenzli\inst{11}
\and
J. Lannier\inst{9}
\and
C. Perrot\inst{4}
\and
S. Peretti\inst{12}
\and
E. Sissa\inst{10}
}
\institute{Max Planck Institute for Astronomy, K\"onigstuhl 17, 69117 Heidelberg, Germany
\and{Instituto de Física y Astronomía, Facultad de Ciencias, Universidad
de Valparaíso, Av. Gran Bretaña 1111, Playa Ancha, Valparaíso,
Chile}
\and{ICM nucleus on protoplanetary disks, ``Protoplanetary discs in ALMA Early Science'', Chile}
\and{LESIA, Observatoire de Paris, PSL Research University, CNRS, Sorbonne Universités, UPMC Univ. Paris 06, Univ. Paris Diderot, Sorbonne Paris Cité, 5 place Jules Janssen, 92195 Meudon, France}
\and{European Southern Observatory, Alonso de Cordova 3107, Vitacura, Santiago, Chile}
\and{Aix Marseille Université, CNRS, LAM (Laboratoire d’Astrophysique de Marseille) UMR 7326, 13388, Marseille, France}
\and{CRAL, UMR 5574, CNRS, Université de Lyon, Ecole Normale Supérieure de Lyon, 46 Allée d’Italie, F-69364 Lyon Cedex 07, France}
\and{Konkoly Observatory, Research Centre for Astronomy and Earth Sciences, Hungarian Academy of Sciences,
P.O. Box 67, 1525 Budapest, Hungary}
\and{Univ. Grenoble Alpes, IPAG, F-38000 Grenoble, France. CNRS, IPAG, F-38000 Grenoble, France}
\and{INAF-Osservatorio Astronomico di Padova, Vicolo dell’Osservatorio
5, 35122, Padova, Italy}
\and{Institute for Astronomy, ETH Zurich, Wolfgang-Pauli-Strasse 27, CH-8093 Zurich, Switzerland}
\and{Geneva Observatory, University of Geneva, Ch. des Maillettes 51, CH-1290 Versoix, Switzerland}}
  \abstract
   {Debris disks correspond to the final evolutionary stage of circumstellar disks around young stars.  Gas-deprived structures seen in debris disks are normally, but not always, attributed to dynamical
   interactions in young planetary systems.}
   {The debris disk of HIP\,73145 has been detected in scattered light in the near-IR, and at far-IR wavelengths before,
but no substructure has been seen so far.  Detection of such substructures in combination
with detailed modeling can hint at the presence of perturbing planetary bodies, or reveal other mechanisms
acting to replenish gas and dust reservoirs and forming structures such as spirals or rings.}
   {We obtained multiwavelength images with SPHERE in the near-IR in the H2 and H3 bands with the IRDIS camera and a 0.95-1.35\,$\mu$m
spectral cube with the IFS. Data were acquired in pupil-tracking mode, thus allowing for angular differential imaging. The SPHERE standard suite of angular differential imaging algorithms was applied. ALMA Band 6 observations complement the SPHERE data.}
   {We detect a bright ring of scattered light plus some secondary structures inside, at least one of them forming a secondary, concentric ring with the first.
   This is the first detection of this disk in total-intensity scattered light.
   A second object is detected in the field at high contrast but concluded to be a background star. Forward modeling yields information on the primary parameters of the disk
 and confirms that the detected substructures are not due to the data analysis approach, which sometimes leads to spurious structures. }
   {We detect a series of concentric rings in the disk around HIP\,73145. This is one of the rare cases where multiple components are necessary to fit the SED and are also detected in scattered light. The presence of such ring structures somewhat questions the nature of the object as a pure debris disk, but the gas and dust content would presumably offer sufficient explanations for such structures to form.}

   \keywords{Stars: individual (HIP 73145) – Protoplanetary disks – Planet-disk interactions – Stars: early-type – Techniques:
image processing – Techniques: high angular resolution}

   \maketitle


\section{Introduction}

\begin{figure*}
\includegraphics[width=2.0\columnwidth]{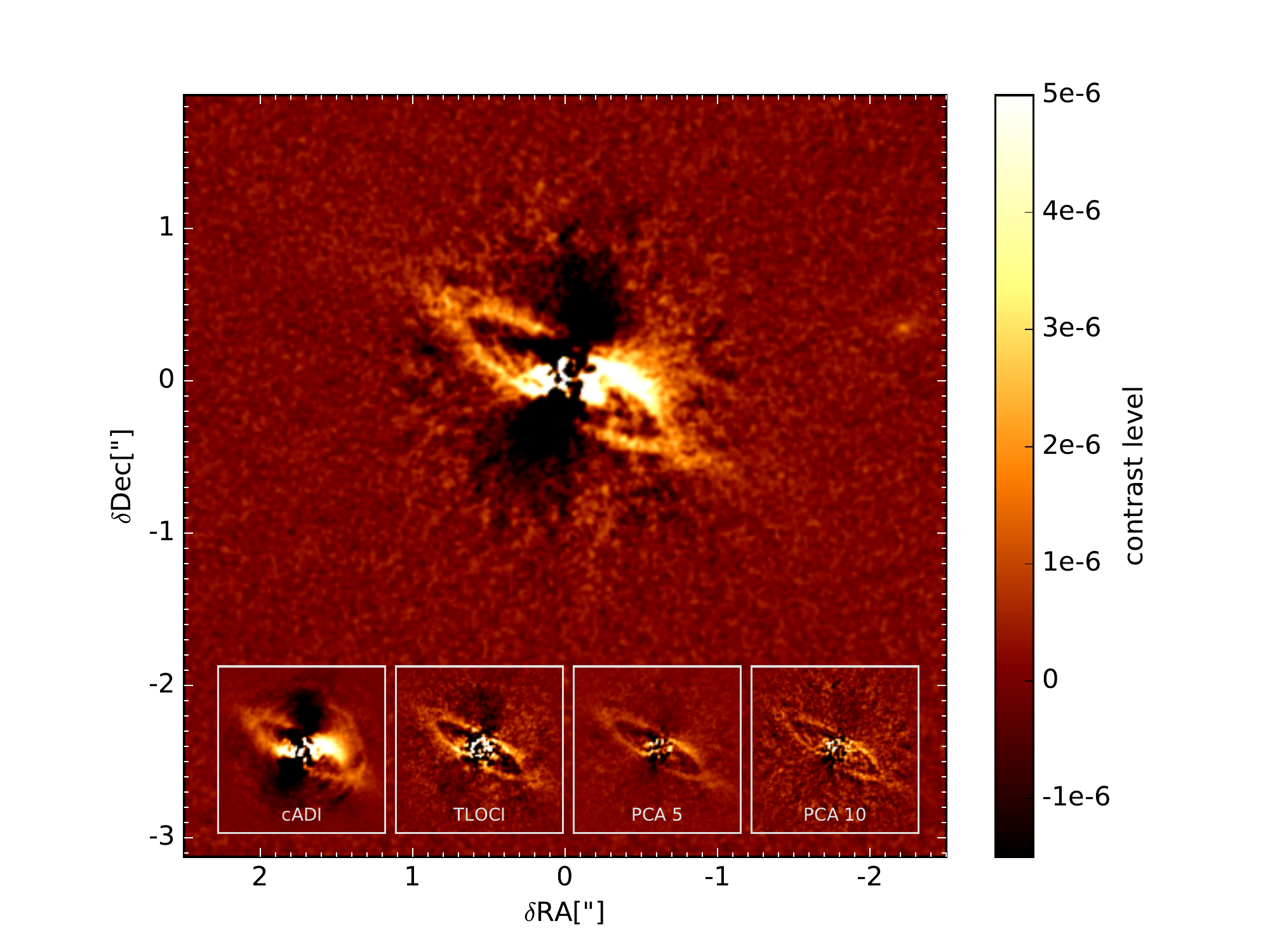}
\caption{HIP\,73145, its disk, and the tentative companion as seen by IRDIS. The large image is a median combination of PCA reductions with 5, 7, 10, and 15 components. The results of other ADI analyses can be seen in the insets. For a description of these results, see text. \label{fig:oview}}
\end{figure*}

\object{HIP\,73145}, also known as \object{HD\,131835}, is an A2IV star (\citealp{Houk1982}) 123\,pc from Earth (\citealp{vanLeeuwen2007}) and has an age of about 15\,Myr (\citealp{Pecaut2012}).
It is a part of the Upper Centaurus Lupus (UCL) moving group (\citealp{Rizzuto2011}), which in turn is part of the Sco-Cen association.
The IR emission of HIP\,73145 was first reported by \citet{Moor2006}  after a search through ISO and IRAS databases, establishing the presence of a debris disk in the system. \citet{Chen2012} presented MIPS observations, which showed $L_{IR}/L_\star = 1.4\times10^{-3}$,
making it one of only four A-type stars in UCL/Lower Centaurus with comparably high excess. \citet{Moor2015} presented Herschel observations of the object.

Debris disks are the successors of gas-rich primordial disks enabling the planet formation process. Once these early disks have been dissipated, dust and gas are continuously produced by collisional interaction or evaporation of the circumstellar bodies formed earlier. Usually, such debris disks exhibit a low gas content. In the case of HIP\,73145, however, \citet{Moor2015} discovered circumstellar gas and provided a lower boundary of $5.2\times10^{-4}\,\mathrm{M}_\oplus$ for the CO gas mass. Newer measurements on optically thin \element[][]{C}\element[][18]{O} indicate a possibly even higher CO gas mass of $0.06\,\mathrm{M}_\oplus$ (\citealp{Moor2016}).  Indications are that the gas is indeed primordial. Also, the dust mass derived by \citet{Moor2015} and \citet{Hung2015b} is between 0.5\,M$_\oplus$  and 0.27\,M$_\oplus$, which would be more consistent with a  disk in transition between protoplanetary and debris disk instead of a pure debris disk. In view of these circumstances, HIP\,73145 at least makes an unusual debris disk.  At such gas-to-dust ratios, the dust dynamics are expected to be governed by gas-dust interactions.

The debris disk of HIP\,73145 was resolved by \citet{Hung2015} in the mid-infrared.  These authors successfully modeled the spectral energy distribution (SED) of the object and their images using a continuous disk component with an inner radius of 35\,AU and an outer radius of 310\,AU, and two ring components, one with a radius of 105\,AU and one with a radius of 220\,AU.  Recently \citet{Hung2015b} reported the discovery of the disk in scattered, polarized $H$-band light, but not in Stokes I or spectral data.
This investigation finds a different morphology from \citet{Hung2015}, fitting the detection between 75\,AU and 120\,AU with an almost flat surface density profile between 75\,AU and 210\,AU. The existence of two components was also confirmed by the SED fitting by \citet{Moor2015}. \citet{Moor2015} also gave an upper limit of $\dot{M}_{acc} < 5\times 10^{-10}$\,M$_\odot$\,yr$^{-1}$ to the accretion rate estimated from the absence of H$_\alpha$ excess emission and $U$ and $B$ photometry.

\begin{figure}
\includegraphics[width=1.0\columnwidth]{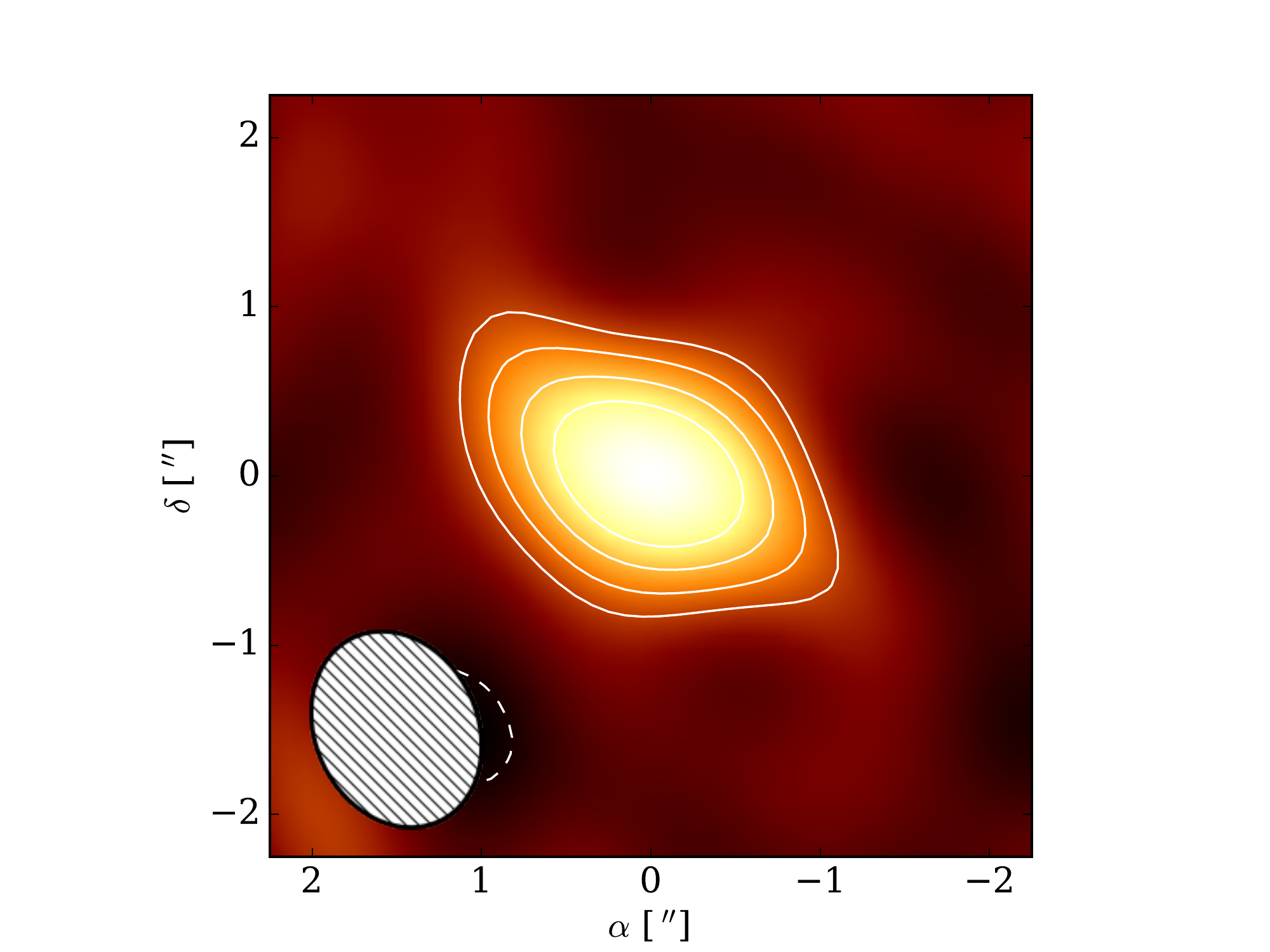}
\caption{Reconstructed ALMA image of HIP\,73145. The linear stretch is between -0.47 and 1.71\,mJy/beam, and the contours are at $[ 3\sigma, 5\sigma, 7.5\sigma, 10\sigma$ with $sigma=0.12$\,mJy/beam.\label{fig:alma-data}}
\end{figure}

Here, we report the results of an observation with SPHERE in nonpolarized scattered light in the $H$ band. We clearly detect the debris disk in the form of a bright ring  with an approximate radius of 96\,AU. A second ring is marginally detected further in, with a radius of 66\,AU.  Tentatively, we also find structures further in at 35\,AU and 45\,AU from the star, but these are very close to the coronagraphic edge and are considered to be less reliable detections. This is the first scattered-light total-intensity detection of the HIP\,73145 disk at near-infrared (NIR) wavelengths.

Additionally we report the results of Band 6 observations with ALMA, where a much more compact disk is detected than in the SPHERE data.

\section{Observation and data reduction \label{sec:obs_red}}
\subsection{SPHERE}

\begin{table}
\caption{SPHERE observations of HIP\,73145 }              
\label{table:obs-sphere}      
\centering                                      
\begin{tabular}{ c c c}          
\hline\hline                        
 Mode/band & DIT $\times$ NDIT $\times$ \#OBs & true north \\    
\hline                                   
    IFS YJ & 64.0s $\times$ 4 $\times$ 16 & -2\degr $\pm$ 0\fdg1\\      
\hline                                             
\end{tabular}
\end{table}

HIP\,73145 was observed with the extreme adaptive optics coronagraphic instrument SPHERE (\citealp{Beuzit2008}, \citealp{Fusco2014}) at the VLT on May 14 2015. The observation was part of the SpHere INfrared survey for Exoplanets (SHINE) campaign aimed at the detection and characterization of extrasolar planets.  HIP\,73145 was observed in the DBI mode of IRDIS (\citealp{Vigan2010}), using the H2 and H3 dual band filters. Simultaneously, a spectral data cube was recorded with the SPHERE integral field spectrograph (IFS; \citealp{Claudi2008}) in YJ mode using 39 spectral channels between 0.95\,$\mu$m and 1.35\,$\mu$m. The observations lasted from 03:30\,UT to 04:41\,UT, resulting in a total field rotation of about 73\degr.

An apodized Lyot coronagraph was used with a mask diameter of 0\farcs185 (\citealp{Boccaletti2008}).
Observing conditions were excellent with a $\tau_0$\footnote{Atmospheric coherence time at 500\,nm.}= 7.2\,ms and seeing between 0\farcs66 and 0\farcs69.  The IRDIS pixel size is $12.25\pm0.02$\,mas resulting in an $11\arcsec \times 12\farcs5$ field of view (FoV).
The IFS has a spatial pixel size of $7.46\pm0.02$\,mas, resulting in a $1\farcs73\times1\farcs73$ FoV.  The observation parameters together with the field orientations derived from astrometric calibrations, as described in  \citet{Maire2016}, are summarized in Tab.~\ref{table:obs-sphere}.

All data were reduced with the SPHERE pipeline (\citealp{Pavlov2008}) and additional analysis tools were implemented at the SPHERE data center (\citealp{SPHERE-DC}).  This includes the usual cosmetics corrections as well as distortion correction and wavelength calibration.
The location of the star is identified by the means of a dedicated star center calibration, where a waffle pattern is applied to the deformable mirror of the adaptive optics system, causing four satellite spots to appear in a subset of images.
In order to remove the stellar halo as far as possible, two different implementations were used: \emph{SpeCal} developed explicitly for the SHINE survey (R. Galicher, priv.~comm.) and another implementation from our team (\citealp{Boccaletti2015}). Both implementations lead to very similar results,
applying a number of angular differential imaging (ADI) techniques: cADI (\citealp{Marois2006}), LOCI (\citealp{Lafreniere2007}), TLOCI (\citealp{Marois2014}), and KLIP (\citealp{Soummer2012}). The results of the data reduction are shown in Fig.~\ref{fig:oview}.

The IFS data do not reveal additional information with respect to the IRDIS data in terms of morphology.  They have only been used to derive improved detection limits in terms of the companion search described in Sec.~\ref{sec:companion-search}.

\begin{figure*}
\includegraphics[width=2.0\columnwidth]{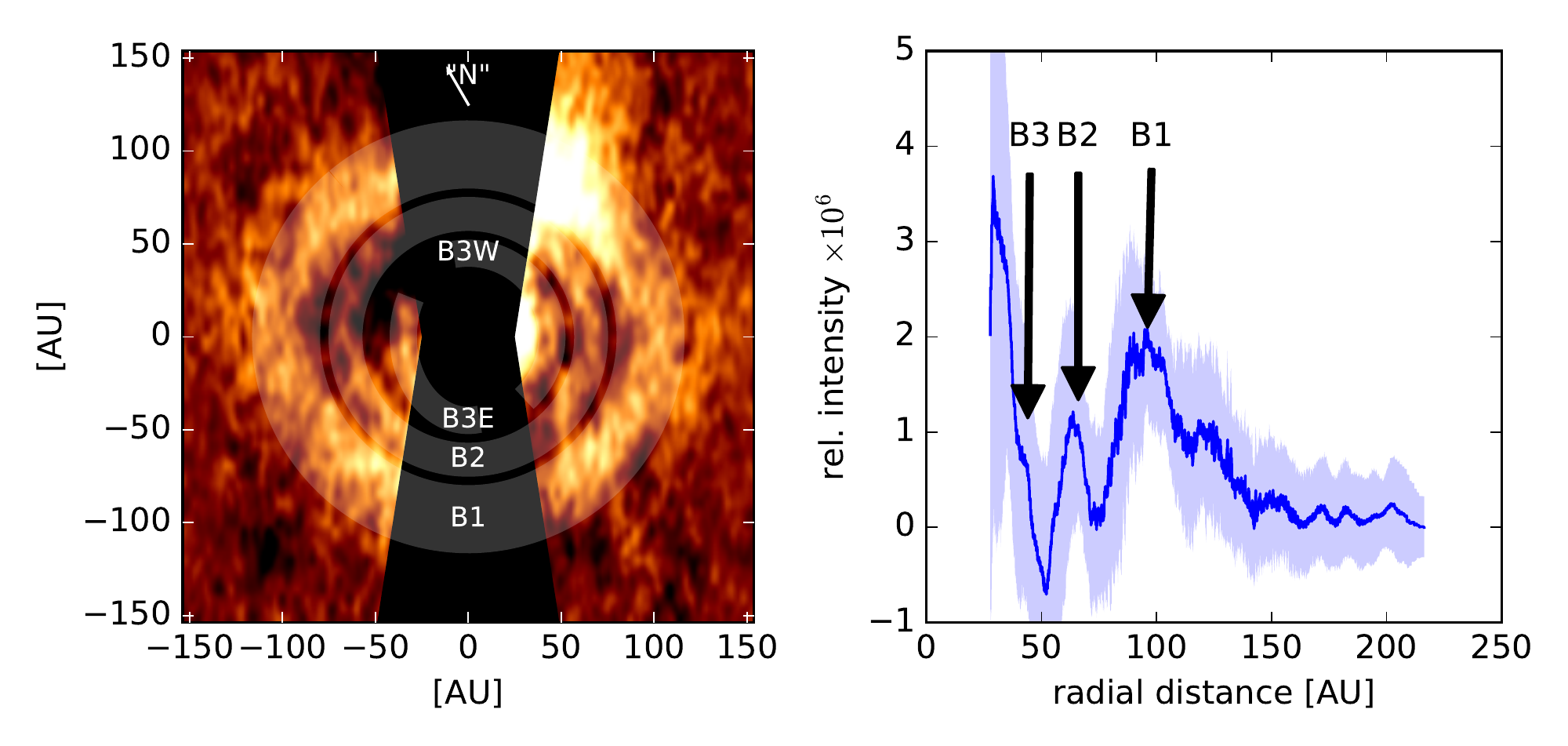}
\caption{Structures detected in the disk of HIP\,73145. Left: Deprojected version of the image shown in Fig.~\ref{fig:oview} (see Text). The features under discussion are denoted with semi-transparent rings and labeled B1, B2, B3E, and B3W.
The black mask covers the area that is too close to the coronagraph and the large negative ADI artifacts. Right: Intensity of all pixels in the image to the left plotted versus their radial separation from the center.  The dark blue line is formed by applying a 100 point wide running mean to all pixel intensities after sorting according to their distance to the center. The light-blue shaded area represents the standard deviation across 100 neighboring points at each separation.
The two major features B1 and B2 are easily identified. The B3 feature is less obvious, as it is close to the inner edge of the usable area and not circular.\label{fig:deproj}}
\end{figure*}

\subsection{ALMA}

On the 10$^{\mathrm{th}}$ of Jan. 2014, HIP\,73145 was observed during Cycle\,0 (program ID 2012.1.00688.S, PI: J.\,Carpenter, Lieman-Sifry et al., submitted) in Band\,6 (245\,GHz, 1.22\,mm).
The precipitable water vapor was 2.7\,mm. We reduced the observations within CASA (\citealp{McMullin2007}) with scripts provided by the observatory.
We exported the complex visibilities from the two broadest spectral windows (128 channels covering 1.875\,GHz each). We averaged the complex visibilities over the 128 spectral channels
and to estimate the uncertainties for each individual baseline, we added in quadrature the standard deviations for the real and imaginary parts of the visibilities.
In Figure\,\ref{fig:alma-data}, we show the reconstructed image obtained with the \texttt{clean} algorithm. The beam has a size of $1.20^{\prime\prime} \times 0.95^{\prime\prime}$, with a position angle of $25.4^{\circ}$, and we measure a standard deviation of $0.12$\,mJy/beam in empty regions of the image.

\section{Morphology of the disk \label{sec:disk-structure}}

When looking closely at Fig.~\ref{fig:oview} it becomes obvious that there are more structures than just a single bright ring with rather sharp outer and inner edges and a semi-major axis of about 96\,AU.  The brightening in the northwestern part of the primary ring appears
independent of the data reduction method (c.f.~Fig.~\ref{fig:oview}), yet it is not absolutely clear whether it is is real or an artifact.

In order to investigate the structures a bit more in detail, we analyzed the deprojected image of the disk assuming a circular ring and an inclination of 72.6\degr from face on. The inclination was found by minimizing the ellipticity of the two ring features described below.
To guide the eye, we denoted apparent azimuthal structures with semi-transparent marks in the left panel of Fig.~\ref{fig:deproj}.  Two of these structures, denoted "B1" and "B2", are also easily found in the plot shown in the right panel of Fig.~\ref{fig:deproj}. This plot shows the pixel intensity versus the radial distance of each pixel from the center. A 100 points-wide running mean has been applied to smooth the line after sorting the pixels into vectors according to their distance from the center. The light blue area denotes the running standard deviation across the 100 pixels neighboring closest in terms of distance from the center.  The B1 and B2 structures were analyzed further by fitting ellipses to the brightest pixels (intensity greater than 1.8$\times 10^{-6}$ for B1 and greater than 1.0$\times 10^{-6}$ for B2) in the areas between the inner and outer edges of the features as taken from Tab.~\ref{table:disk-features}. These fits yield a minimum ellipticity of 0.04 for B1 and 0.08 for B2 for an inclination of 72.6$\pm$1.0\degr. Also the two features appear very well concentric, the center of B2 being offset from that of B1 by 3.3\,AU (0\farcs03) or 3.3\% of the radius of B1. There is a third inner feature, which does not appear to be circular and is denoted as B3E for its eastern and B3W for its western part. The western part has an apparent radius of 45\,AU, whereas the eastern part appears more elliptical; the ellipse denotes the semi-transparent mark in Fig.~\ref{fig:deproj} having a semi-major axis of 45\,AU and a semi-minor axis of 35\,AU. Tab.~\ref{table:disk-features} summarizes the location of the features found in the disk of HIP\,73145. To determine the inner and outer edges given in Tab.~\ref{table:disk-features}, we smoothed the dark blue line in Fig.~\ref{fig:deproj} with a running mean of 1000 points width, roughly (though not constantly) corresponding to about 7\,AU in the distance-sorted vector of all pixels, and read the locations of the zeros of the second derivative closest to the peak positions.  This failed for the inner edges of feature B3 and the outer edge of feature B1, which have been determined by eye.  We stress again that the ADI processing increasingly affects the features inward of B1.

In Sec.~\ref{sec:dr-b2b3}, we present an indication that the ring structures we see after the ADI processing is perfectly consistent with a true ring structure of the object. However, there may still be additional, potentially more extended components present in the object, to which our sensitivity would be greatly reduced by the ADI processing. \citet{Milli2012} investigated the impact of different types of ADI processing on the structural appearance of circumstellar disks and showed that indeed extended structures can be turned into higher frequency features, although ring-like features were created in no case  that had not been present before ADI processing.  Especially for B3, which is not clearly resolved as a ring structure and heavily affected by the stellar core, a smooth, extended inner component could potentially be the cause for what we see.  Summarizing this sanity check, the forward modeling ensures that a ring-like structure would be consistent with the data, we cannot fully exclude that additional components could be present that also might be consistent with the data after ADI processing.

\begin{table}
\caption{Ring features in the disk}              
\label{table:disk-features}      
\centering                                      
\begin{tabular}{ c c c c}          
\hline\hline                        
 ID & peak radius  & inner edge & outer edge \\    
\hline                                   
          &                    &                    & to 140\,AU / 1\farcs14 \\
    B2    & 66.0\,AU / 0\farcs54 & 57\,AU / 0\farcs46 &  71\,AU / 0\farcs58 \\
    B3E   & 35\,AU / 0\farcs28 & -- & 46\,AU / 0\farcs37 \\
    B3W   & 45\,AU / 0\farcs37 & -- & 46\,AU / 0\farcs37 \\  
\hline                                             
\end{tabular}
\end{table}

Such structures in circumstellar disks have been seen around numerous young stars and have sparked a remarkable debate, offering a large set of possible explanations.  Most require a substantial amount of dust, the presence of which in the HIP\,73145 disk is not firmly established to date.
If the speculations triggered by \citet{Moor2016} that the amount of gas in the HIP\,73145 disk is comparable to transitional disks prove to be true,  some of the following ideas may apply:  Planetary companions can cause ring-like structures according to \citet{Picogna2015}, \citet{Dipierro2015}, and \citet{Gonzalez2015}. Not resorting to planets \citet{Flock2015} simulate magnetized disks to form gaps and rings, \citet{Zhang2015} speculate about fast pebble growth near condensation fronts, and \citet{Okuzumi2016} talk about sintering-induced dust rings.  See \citet{Okuzumi2016} also for an overview of even more possible explanations for concentric ring structures to form in circumstellar disks. In case the dust content in HIP\,73145 is as low as expected from the classification as a debris disk, \citet{Lyra2013} offer an explanation of how to form sharp rings in planet-devoid gas disks with a gas-to-dust ratio around unity as may be the case here. They found a robust clumping instability caused by dust-gas interactions that organized the dust into narrow concentric rings.  Also \citet{Klahr2005} already described a similar mechanism to form rings of dust in debris disks without the presence of planets. In their case, the clumping instability arises in regions where an enhancement in the dust density leads to local gas temperature and pressure increase, concentrating the dust into rings. These authors predict such patterns also remain after the gas has been removed from the disk.

Our findings of the concentric rings can now be used to confront the various models described above, but this is beyond the scope of this paper.  The presence of a possible planetary companion in the stellar vicinity is explored in the next section by considering the combined IFS and IRDIS detection limits.

\begin{figure*}
\includegraphics[width=1.0\columnwidth]{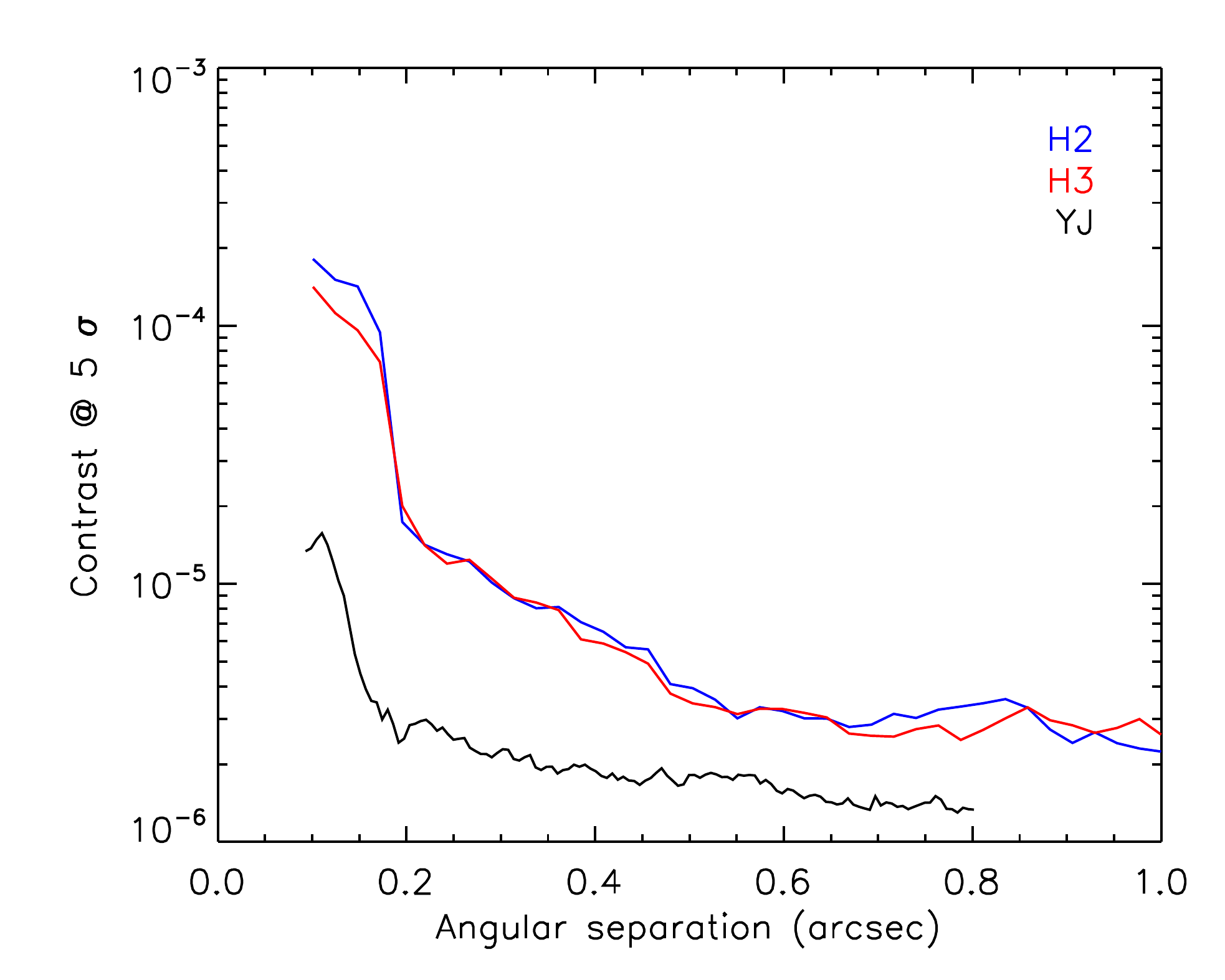}
\includegraphics[width=1.0\columnwidth]{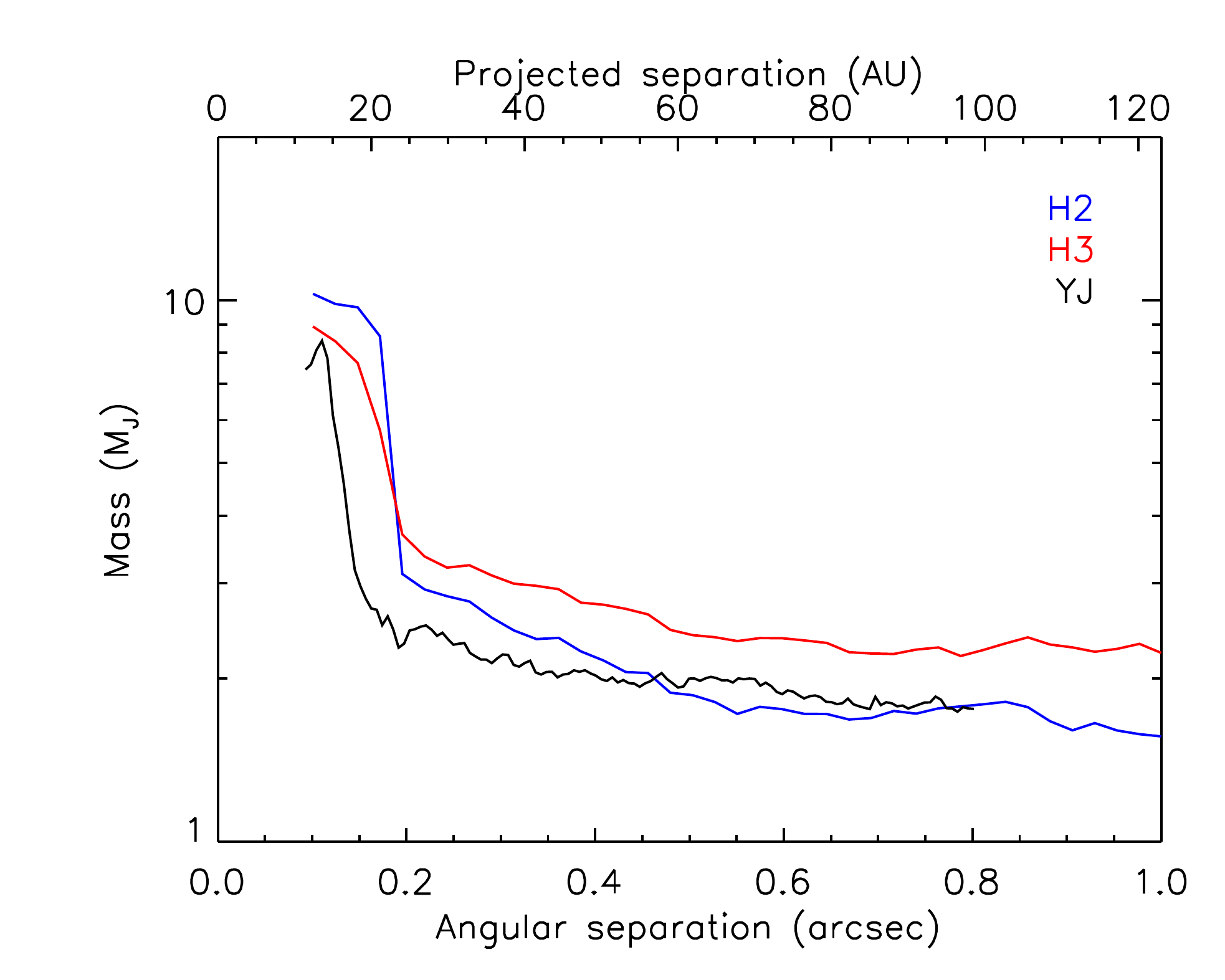}

\caption{Contrast achieved in the companion search with IRDIS (H2 and H3) and IFS (YJ) using the TLOCI algorithm. The ADI contrast limits from the left figure have been converted to mass limits on the right using models from \citet{Baraffe2015}
computed specifically for the SPHERE spectral bands and an age of 10\,Myr\label{fig:companion-search}.}
\end{figure*}

\section{Companion search \label{sec:companion-search}}

A suite of algorithms was run on both the IRDIS and IFS data to detect possible companions to HIP\,73145 (see Sec.~\ref{sec:obs_red}).  The resulting contrast limits are shown in the left plot in Fig.~\ref{fig:companion-search}.
The raw ADI contrast limits were converted to mass limits using the atmospheric model by \citet{Baraffe2015} computed specifically for the SPHERE filter set (among others), and assuming an age of 10\,Myr via the COND evolutionary tracks by \citet{Baraffe2003}. We can exclude companions
between 0\farcs2 and 1\arcsec with a flux of more than $3\times10^{-6}$ times the stellar flux (or a contrast of less than 13.8\,mag) in YJ, more than $2\times10^{-5}$ times the stellar flux in H2 and H3, or a mass of more than $\propto 2.5\,M_\mathrm{Jup}$.

\begin{figure}
\includegraphics[width=1.0\columnwidth]{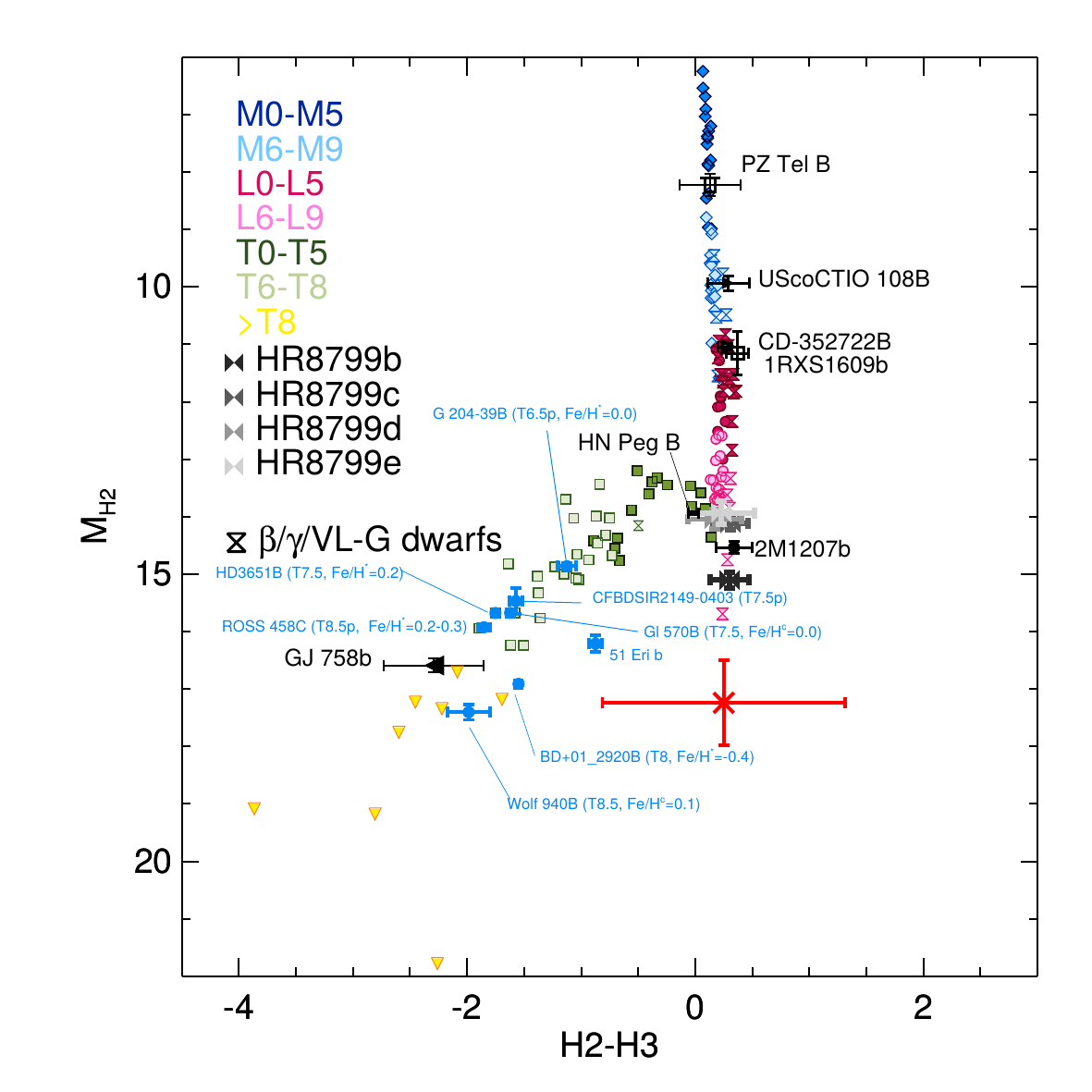}

\caption{Color-magnitude plot for the companion found near HIP\,73145 together with locations of synthetic SPHERE photometry of low gravity ($\beta$/$\gamma$/VL-G) dwarfs and old field MLTY dwarfs. \label{fig:companion-found}}
\end{figure}

One apparent companion was found however at a separation of 2180\,mas$\pm$ 25\,mas, a position angle of 278\fdg4 $\pm$ 0\fdg8, and a contrast of 15.12\,mag$\pm 0.7$\,mag in H2 and 14.87\,mag $\pm 0.8$\,mag in H3. This object is clearly visible in Fig.~\ref{fig:oview}.
Its magnitude is right at the contrast limit. The color-magnitude plot for the object is shown in Fig.~\ref{fig:companion-found}, together with synthetic SPHERE photometry of low gravity ($\beta$/$\gamma$/VL-G) dwarfs and old field MLTY dwarfs. This photometry is derived from the flux-calibrated near-infrared spectra of the sources gathered from the SpeXPrism library \citep{2014ASInC..11....7B}, and additional studies \citep{2010A&A...517A..76P, 2013ApJ...772...79A, 2014A&A...562A.127B, 2010ApJ...725.1405B, 2011ApJ...729..139W, 2015ApJ...804...96G, 2015ApJ...804...92S, 2014AJ....147...34S, 2015ApJ...799..203G, 2013ApJS..205....6M, 2013ApJ...777L..20L, 2010ApJ...719..497L, 2008A&A...482..961D}. We added the photometry of benchmark T-type companions (Bonnefoy et al. in prep, for the details) identified around nearby stars and one of the exoplanet 51 Eri b derived from the Gemini Planet Imager (GPI) spectrum of the object \citep{2015Sci...350...64M}. At its negligible H2-H3 color, the object appears way too faint to belong to the sequence of comparable substellar objects. We thus suspect that this must be a background object.

\section{Modeling of the SPHERE/IRDIS scattered light images}\label{sec:disk_modeling}

The debris disk around HIP\,73145 has already been resolved in the near- and mid-IR (\citealp{Hung2015,Hung2015b}), but with the new VLT/SPHERE observations, we are able to resolve details in the structure of the debris disk that have not been seen before. We are therefore able to provide new and valuable constraints on the morphology of the disk by following a forward modeling approach that can account for the effects of angular differential imaging (e.g., self subtraction; \citealp{Milli2012}). However, given that both the B3 and B2 belts are much fainter than the B1 disk, and that B3 is located close to the coronagraph, we only attempt to model the outermost B1 disk. We adopt the same modeling strategy as presented in \citet{Olofsson2016}, which we briefly summarize in the following.

\subsection{Stellar parameters}

The photospheric model for HIP\,73145 is taken from the ATLAS9 Kurucz library (\citealp{Castelli1997}) for an effective temperature of 9000\,K (\citealp{Wright2003}) and log$g$ of $4.0$ (\citealp{Chen2014}).
At a distance of $122.7^{+16.2}_{-12.8}$\,pc (\citealp{vanLeeuwen2007}), using the dilution factor to scale the photospheric model to the optical and 2MASS observations, we find a radius of $1.38$\,R$_{\odot}$
and a luminosity of $11.5$\,L$_{\odot}$. While fitting the photospheric model to the observations, we also include foreground extinction and find a value of $A_{\mathrm{V}} = 0.36$\,mag. For the stellar mass,
which only matters when discussing radiation pressure in Section\,\ref{sec:sed}, we assume the same value as in \citet[][$1.77$\,M$_{\odot}$]{Moor2015} even though they derived a slightly different effective
temperature and obtained a lower luminosity (using log$g$\,$ = 4.25$).

\subsection{Modeling strategy}

We consider seven free parameters. The radial dust density distribution is parametrized by the reference radius $r_0$, and the inner and outer power-law slopes ($\alpha_{\mathrm{in}}$ and $\alpha_{\mathrm{out}}$, respectively). The density distribution is given by
\begin{equation}\label{eqn:dens}
n(r) \propto \left[ \left( \frac{r}{r_0} \right)^{-2\alpha_{\mathrm{in}}} + \left( \frac{r}{r_0} \right)^{-2\alpha_{\mathrm{out}}} \right] ^{-1/2}.
\end{equation}
The inclination $i$ and position angle $\phi$ of the disk are also free parameters in the modeling. For $i = 0^{\circ}$ the disk is seen face-on, and for $\phi = 0^{\circ}$ the semi-major axis of the disk is along the north-south axis, with the front side of the disk toward the east. Increasing $\phi$ rotates the disk counter-clockwise. We include the opening angle $\psi$ of the disk in the pool of free parameters in an attempt to constrain its vertical height. The vertical dust distribution follows a Gaussian profile with a $\sigma$ width of $\psi$. To easily try to reproduce the phase function of the scattered light, we opt for the Henyey-Greenstein (HG) analytical prescription (\citealp{Henyey1941}), controlled by the asymmetry parameter $g$.

For a given set of free parameters, we produce a synthetic image using the \texttt{DDiT} code described in \citet{Olofsson2016}. We multiply the image by the last free parameter $f_{\mathrm{ADI}}$ (a scaling factor to try to match the intensity of the observations) and generate a cube of disk images rotated to match the parallactic angles of the SPHERE observations. We then subtract the synthetic cube to the observed cube and run the principal component analysis (PCA; using the implementation of the \texttt{scikit-learn} Python package, \citealp{Pedregosa2011}), keeping the first five components. The goodness of fit is evaluated as the sum over all pixels of the PCA output to the square. In order to decrease the computational time, we downsampled and resized the observed datacube. The images are $200 \times 200$ pixels in size and the pixel scale is twice as large as the original observations ($0.0245^{\prime\prime}$). A numerical central mask of radius $0.15^{\prime\prime}$ is applied to both the model and observations. To find the best-fit model, we use an affine invariant ensemble sampler Monte Carlo Markov Chain, implemented in the \texttt{emcee} Python package (\citealp{Foreman-Mackey2012}). We use 100 ``walkers'', a burn-in phase of 1\,000 runs and then run the chains over 1\,000 more iterations. The mean acceptance fraction over all walkers is 0.40 (within the recommended boundaries of the documentation of \texttt{emcee}), and the maximum autocorrelation time of 70 steps.

\subsection{Modeling results}

\begin{figure*}
\includegraphics[width=2.0\columnwidth]{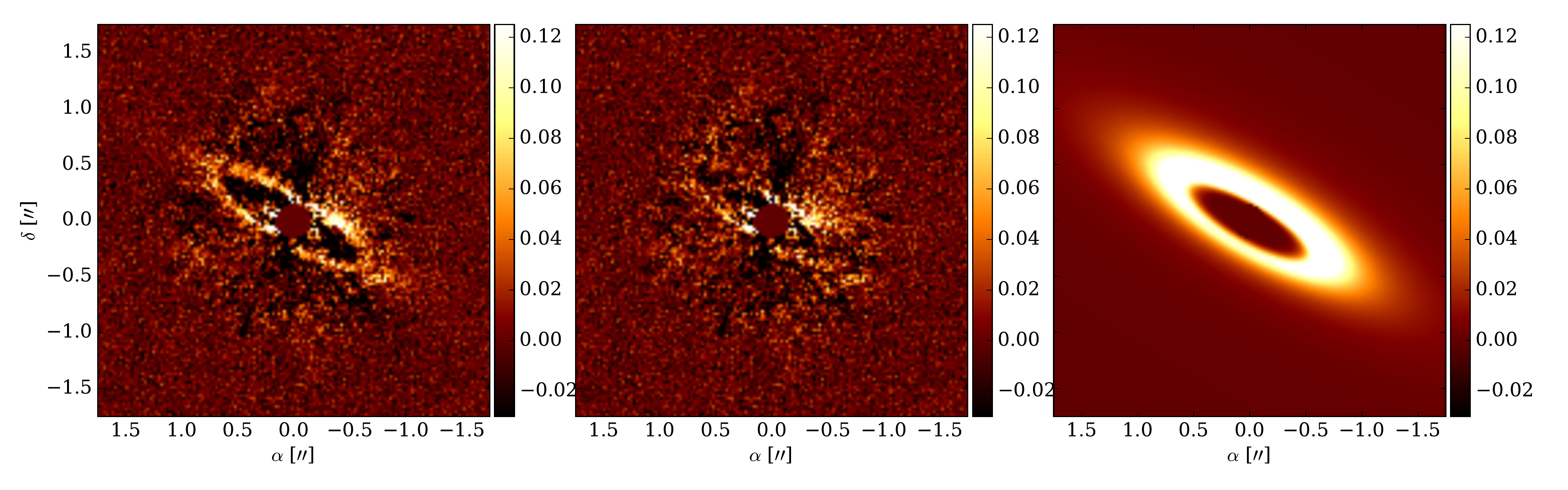}
\caption{Observations, residuals, and best-fitting model (left to right, respectively) for the IRDIS $H$-band observations of HIP\,73145. We removed the 5 main components from the PCA.}
\label{fig:disk_results}
\end{figure*}

\begin{figure*}
\includegraphics[width=2.0\columnwidth]{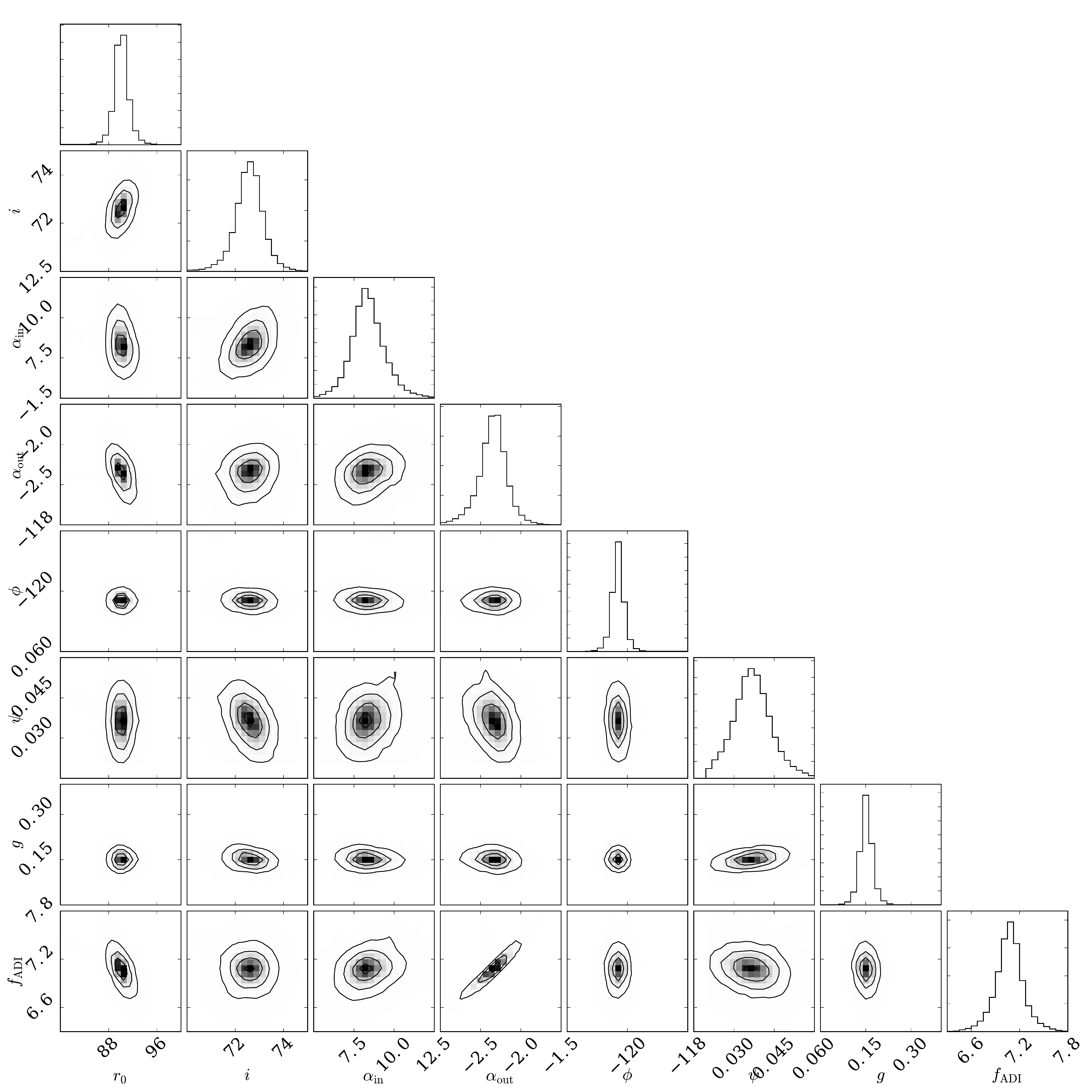}
\caption{Projected probability distributions for the modeling of the SPHERE/IRDIS observations.}
\label{fig:disk_triangle}
\end{figure*}

\begin{table}
\centering
\caption{Best-fit results for the modeling of the SPHERE observations.\label{tab:disk_sphere}}
\begin{tabular}{@{}lccc@{}}
\hline\hline
Parameter & Uniform prior & $\sigma_{\mathrm{kde}}$ & Best-fit value \\
\hline
$r_0$ [au]              & $[70, 150]$     & $0.5$   & $90.1_{-1.1}^{+1.2}$ \\
$i$ [$^{\circ}$]        & $[60, 88]$      & $0.1$   & $72.6_{-0.6}^{+0.5}$ \\
$\alpha_{\mathrm{in}}$  & $[2.5, 15]$     & $0.1$   & $8.2_{-0.8}^{+1.1}$ \\
$\alpha_{\mathrm{out}}$ & $[-10, -1.5]$   & $0.1$   & $-2.3_{-0.2}^{+0.2}$ \\
$\phi$ [$^{\circ}$]     & $[-140, -100]$  & $0.1$   & $-120.3_{-0.2}^{+0.2}$ \\
$\psi$ [$^{\circ}$]     & $[0.02, 0.12]$  & $0.005$ & $0.04_{-0.01}^{+0.01}$ \\
$g$                     & $[-0.95, 0.95]$ & $0.01$  & $0.15_{-0.02}^{+0.02}$ \\
$f_{\mathrm{ADI}}$      & $[1, 10]$       & $0.1$   & $7.1_{-0.2}^{+0.2}$  \\
\hline
\end{tabular}
\end{table}

Figure\,\ref{fig:disk_results} shows the observations, the residuals, and the best-fit model we obtain. Most of the disk is removed, except for what appears to be azimuthal variations close to the coronagraph. As can be seen in Figure\,\ref{fig:disk_triangle}, which shows the projected probability distributions across the parameter space, all parameters are well constrained. Table\,\ref{tab:disk_sphere} summarizes our best-fit results, and the uncertainties are derived using a kernel density estimator on the probability distributions (the width of the Gaussian kernel is reported in Tab.\,\ref{tab:disk_sphere}; see \citealp{Olofsson2016} for more detail).

We find that the dust density distribution peaks at about 90\,au from the star, with a steep inner slope ($\alpha_{\mathrm{in}} \sim 8$) and a shallow outer slope ($\alpha_{\mathrm{out}} \sim -2.5$). The disk has an inclination of $72.6^{\circ}$, and with a position angle of $-120^{\circ}$, the front side of the disk would be toward the northwest corner. One should note that the location of the front side (closer to the observer) is ambiguous. The northwest side seems a little brighter and we find a mildly asymmetric value for $g$ (0.1). Given that phase functions for debris disks are often extremely peaked for small scattering angles (e.g., Milli et al. submitted, \citealp{Olofsson2016}), we hypothesize that the front side of the disk is toward the northwest side, but this result remains to be confirmed (especially if the bright spot is an artefact, caused by wind during the observations, for instance).

\citet{Moor2015} found that the debris disk was marginally resolved with the \textit{Herschel}/PACS observations at 70 and 100\,$\mu$m. They used observations of $\alpha$\,Boo, $\alpha$\,Tau, $\alpha$\,Cet, and $\beta$\,And to derive a reference PSF and found that the observations of HIP\,73145 were slightly wider than the reference PSF along the semi-major axis. We computed images of our best-fit model at 70 and 100\,$\mu$m and convolved these images with PSFs with FWHM of 5.7$^{\prime\prime}$ and 6.8$^{\prime\prime}$, respectively, comparable to the beam sizes of the \textit{Herschel} observations. We measured the profile along the semi-major and semi-minor axes of the disk and obtained FWHMs of 6\farcs6 $\times$ 5\farcs8 at 70\,$\mu$m, and 7\farcs6 $\times$ 6\farcs9 at 100\,$\mu$m. These values are within the uncertainties reported by \citet{Moor2015}, suggesting our best-fit model is compatible with the \textit{Herschel}/PACS observations.

\section{Modeling of the ALMA observations}\label{sec:alma}

\begin{figure*}
\includegraphics[width=2.0\columnwidth]{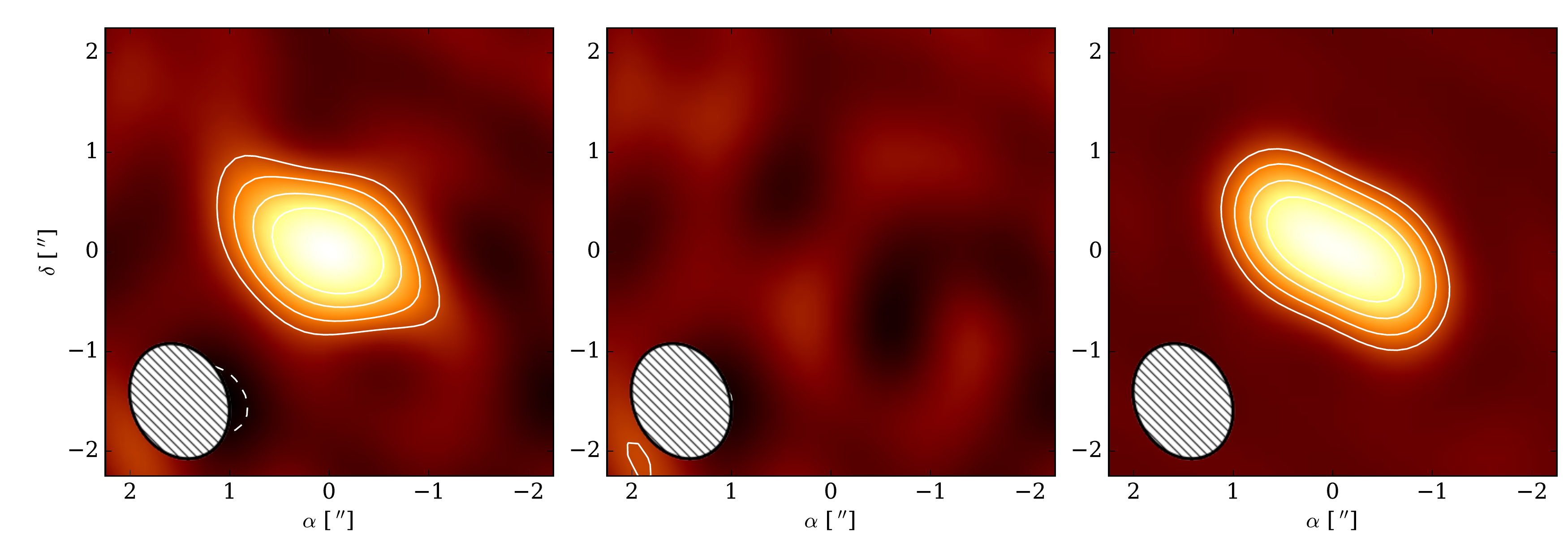}
\caption{Observations, residuals, and best-fit model (convolved through CASA) for the Band\,6 ALMA data (from left to right, respectively). Each panel has the same linear stretch between $-0.47$ and $1.71$\,mJy/beam, and the contours are at $[3\sigma$, $5\sigma$, $7.5\sigma$, $10\sigma$$]$, with $\sigma = 0.12$\,mJy/beam.
}
\label{fig:alma}
\end{figure*}

\begin{figure}
\includegraphics[width=1.0\columnwidth]{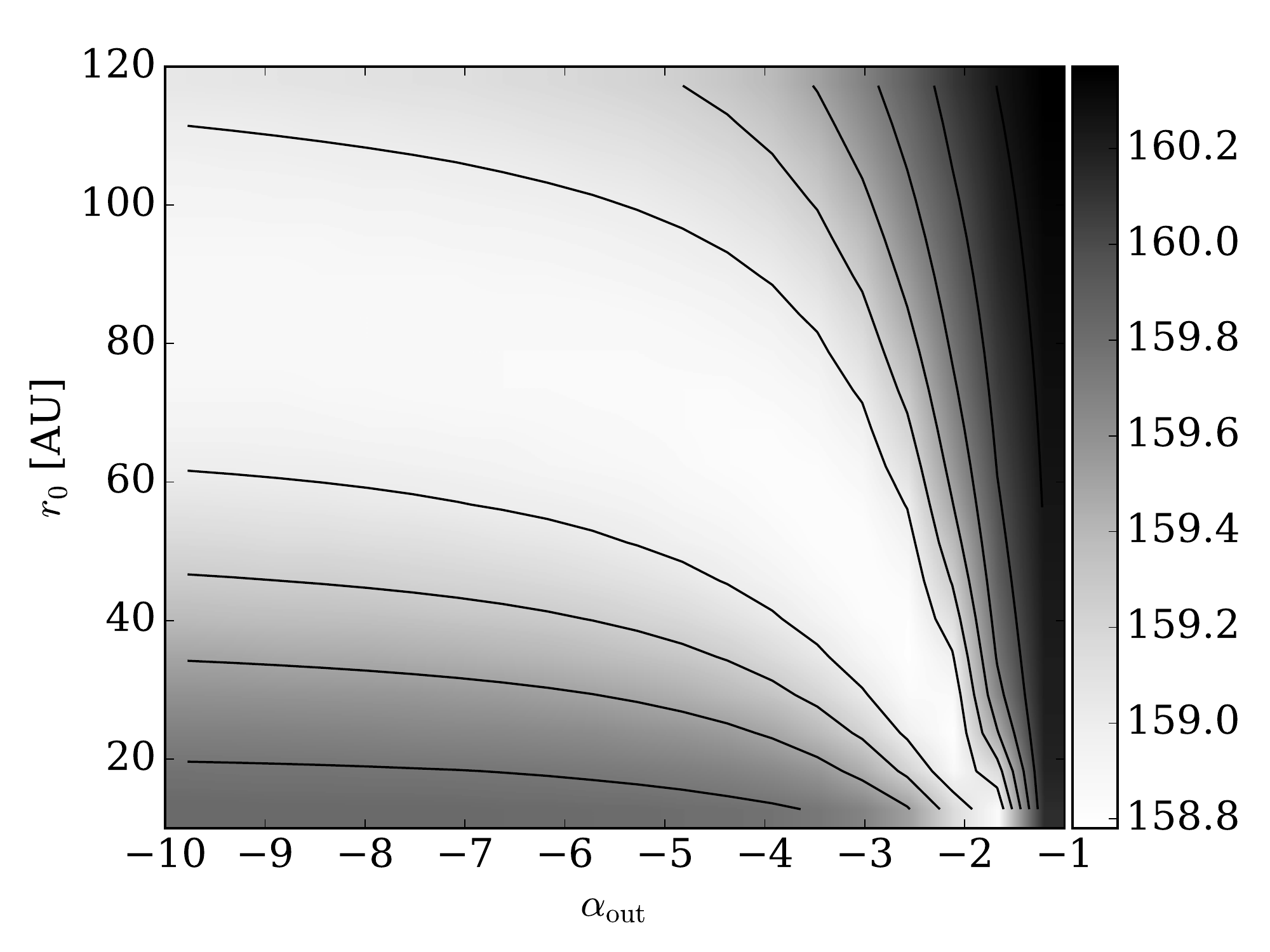}
\caption{Cut of the $\chi^2$ cube for the best value of the Band\,6 flux ($f_{\mathrm{mm}} = 3.5$\,mJy) along the dimensions of $r_0$ and $\alpha_{\mathrm{out}}$.}
\label{fig:alma_chi2}
\end{figure}

To model the ALMA observations, we proceeded as in \citet{Olofsson2016}: for a given set of parameters, we compute a synthetic image at 1.22\,mm, scale the total flux to the free parameter $f_{\mathrm{mm}}$ (in mJy), compute the Fourier transform of the image, and interpolate it at the proper spatial frequencies to compute the complex visibilities. Given that the disk is barely resolved (see Fig.\ref{fig:alma}), performing a detailed modeling with many free parameters may lead to unreliable results. Furthermore, because of the large beam, there appears to be a degeneracy between the reference radius of the disk $r_0$ and the outer slope of the density distribution $\alpha_{\mathrm{out}}$. Consequently, the MCMC approach we used for the SPHERE data has difficulty providing meaningful constraints on the disk parameters, and we instead perform a simple $\chi^2$ grid search over the following three free parameters: the reference radius $r_0$, the outer slope of the density distribution $\alpha_{\mathrm{out}}$, and the mm flux $f_{\mathrm{mm}}$. All other parameters (e.g., inclination and position angle) are the best-fit results from the modeling of the SPHERE observations. We do not include the central star in the synthetic image. Inspecting the real part of the visibilities at the shortest baselines, we can expect a total flux of a few mJy, while the star is expected to have a flux that is close to a few $\mu$Jy at the wavelength of the observations.

In Figure\,\ref{fig:alma} we show the observations, residuals, and best-fit model obtained for $r_0 = 21.6$\,au, $\alpha_{\mathrm{out}} = -1.95$, and $f_{\mathrm{mm}} = 3.5$\,mJy (from left to right, respectively). Each panel has the same linear stretch (from $-0.47$ to $1.71$\,mJy/beam) and the contours show the $[3\sigma, 5\sigma, 7.5\sigma, 10\sigma]$ levels, with $\sigma = 0.12$\,mJy/beam (estimated in empty regions of the observations). It seems at first that the disk as seen with ALMA is much more compact than the disk we detected with SPHERE. However, we find that the observations are best reproduced by a family of models as there is a significant degeneracy between the reference radius of the disk and the outer slope of the density distribution. For this family of models, we find that the flux at 1.22\,mm is of about $f_{\mathrm{mm}} = 3.5$\,mJy, even though models with large $r_0$ seem to require slightly larger $f_{\mathrm{mm}}$ ($\sim 4$\,mJy). Figure\,\ref{fig:alma_chi2} shows a cut of the 3D $\chi^2$ cube (for the best-fit value of $f_{\mathrm{mm}} = 3.5$\,mJy). One can clearly see that there is a valley of low $\chi^2$ values; as the reference radius increases, the slope for the density distribution steepens. Therefore, with the available observations, we can only provide conservative confidence intervals for $r_0$ ($[20,110]$\,au) and $\alpha_{\mathrm{out}}$ ($[-10.0, -2.0]$).

If one assumes that indeed $r_0 = 90$\,au, the differences for $\alpha_{\mathrm{out}}$ between the modeling results of the SPHERE and the ALMA observations can be reconciled by considering the effect of radiation pressure. The ALMA observations trace the population of (sub-)\,mm-sized grains, while the scattered light observations trace a population of smaller grains ($\mu$m-sized) that are subject to efficient radiation pressure around an A-type star. Therefore the disk would appear more extended in the SPHERE observations than in the ALMA observations.  We cannot further address these differences with the available dataset and observations at higher angular resolution in the (sub-)mm are required.

\section{Modeling the SED}\label{sec:sed}

The SED of HIP\,73145 shows a strong excess that already starts to be significant in the mid-IR.
Several studies reported that its SED is best reproduced using Planck functions at different temperatures (e.g., \citealp{Moor2015}).
Here we attempt to reproduce the observed SED (constructed with the fluxes reported in \citealp{Moor2015}) and to constrain the properties of the dust grains
(grain size distribution and dust composition). We follow the same procedure as described in \citet{Olofsson2016}: we used absorption coefficients
calculated with the Mie theory, between $0.01$\,$\mu$m and 5\,mm. For the optical constants, we consider amorphous silicate grains with olivine stoichiometry
(\citealp{Dorschner1995}, $\rho = 3.71$\,g.cm$^{-3}$), amorphous water ices (\citealp{Li1998}, $\rho = 1.2$\,g.cm$^{-3}$), and amorphous carbon (\citealp{Jager1998},
$\rho = 1.67$\,g.cm$^{-3}$). We mix the optical constants for fractions of amorphous carbon, water ices, and porosity in increments of 10\%, using the Bruggeman mixing theory.
We thus create a library of 1000 absorption efficiencies files that are used during the modeling. With the SPHERE observations, we are able to constrain the reference radius of
the disk and we thus keep it fixed when modeling the SED. The free parameters are the outer slope of the dust density distribution ($\alpha_{\mathrm{out}}$), grain size
distribution slope $p$ (d$n(s) \propto s^p$d$s$, $p \leq 0$), minimum grain size $s_{\mathrm{min}}$ ($s_{\mathrm{max}}$ is fixed at $5$\,mm), fraction of amorphous water
ices, carbon, and porosity. Preliminary attempts to fit the SED with a single dust belt failed to reproduce the shape in the mid-IR, therefore we include a second reference radius
$r_{\mathrm{inner}}$ for an inner belt; we discuss the necessity of including a third belt later on. For a given set of parameters, we fit the SED as follows:
we compute the thermal emission arising from the main belt at $r_0$, and save the emission profile (regardless of the dust mass). We then compute the emission from the inner
belt at $r_{\mathrm{inner}}$, using the same dust composition, size distribution, and density distribution. We also save the emission profile as a function of the wavelength.
We fit the observed fluxes with a linear combination of the two emission profiles using a Levenberg-Marquardt solver. Since the emission is optically thin, the scaling factors are used to estimate the dust masses in the inner and outer belts. We then compute the $\chi^2$ (using the weighting scheme described in \citealp{Olofsson2016}), which is fed to the \texttt{emcee} sampler. We use 50 walkers, a burn in sequence of 500 steps and we then run the MCMC for 1000 steps in total. At the end, we find an acceptance fraction of $0.21$ and a maximum autocorrelation time of $89$ steps.

\subsection{SED modeling results \label{sec:sed-modeling-results}}

\begin{table}
\centering
\caption{Best-fit results for the modeling of the SED.\label{tab:sed}}
\begin{tabular}{@{}lccc@{}}
\hline\hline
Parameter & Uniform prior & $\sigma_{\mathrm{kde}}$ & Best-fit value \\
\hline
$\alpha_{\mathrm{out}}$          & $[-5, -0.5]$  & $0.1$   & $-3.9_{-0.4}^{+0.3}$ \\
$p$                              & $[-5, -2]$    & $0.02$  & $-3.41_{-0.04}^{+0.05}$ \\
$s_{\mathrm{min}}$ [$\mu$m]      & $[0.01, 20]$  & $0.1$   & $8.1_{-0.2}^{+0.5}$ \\
$f_{\mathrm{ice}}$               & $[0, 0.95]$   & $0.1$   & $0.1_{-0.1}^{+0.1}$ \\
$f_{\mathrm{carbon}}$            & $[0, 0.95]$   & $0.1$   & $0.8_{-0.1}^{+0.1}$ \\
$f_{\mathrm{porosity}}$          & $[0, 0.95]$   & $0.1$   & $0.2_{-0.1}^{+0.2}$ \\
$r_{\mathrm{inner}}$ [au]        & $[0.5, 15]$   & $0.1$   & $5.4_{-0.3}^{+0.3}$ \\
M$_{\mathrm{inner}}$ [M$_{\oplus}$] &               & $0.005$ & $0.025_{-0.008}^{+0.011}$ \\
M$_{\mathrm{outer}}$ [M$_{\oplus}$] &               & $0.5$   & $7.2_{-0.6}^{+0.6}$ \\
\hline
\end{tabular}
\end{table}

\begin{figure}
\includegraphics[width=1.0\columnwidth]{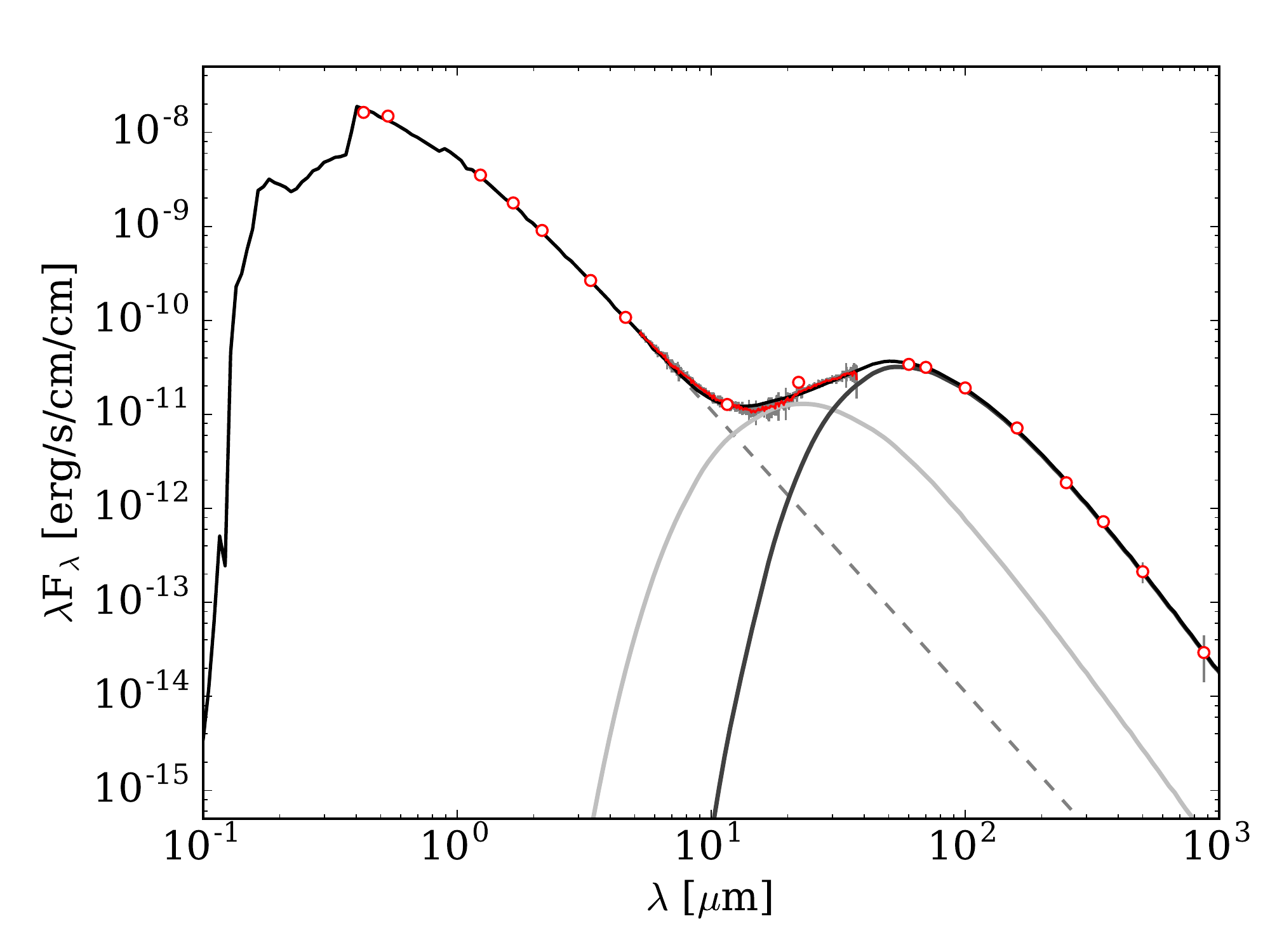}
\caption{Debris disk SED around HIP\,73145. Observations are shown in red; the photospheric model is shown with a dashed gray line. The emission profiles of the inner and outer belts are shown in light and dark gray lines, respectively, while the total fit (including the stellar contribution) is indicated in solid black.}
\label{fig:sed}
\end{figure}

\begin{figure*}
\includegraphics[width=2.0\columnwidth]{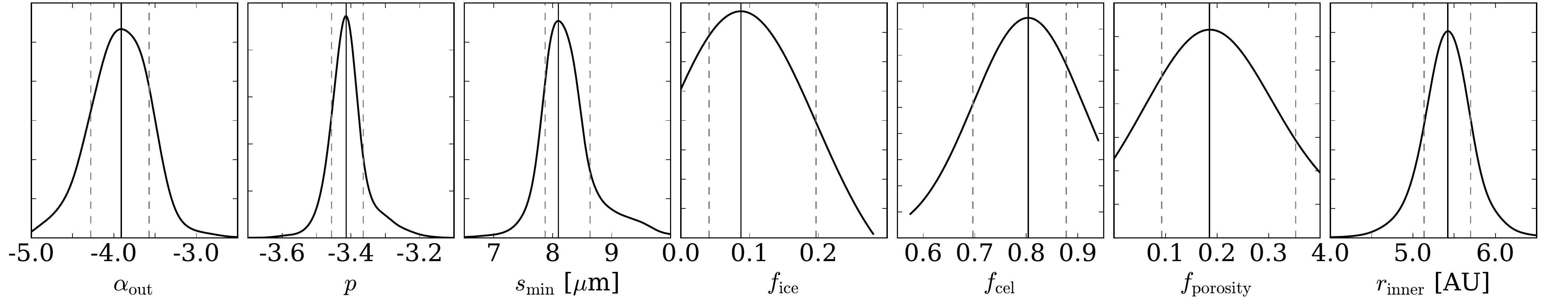}
\caption{Projected probability distributions for the results of the SED modeling.}
\label{fig:sed_proba}
\end{figure*}

Figure\,\ref{fig:sed} shows the observed and modeled SED of the debris disk around HIP\,73145. The observations are shown in red, the stellar model in dashed gray, and the total best-fit model in solid black. The thermal emission contributions of the inner and outer belts are shown in light and dark grays, respectively. One can see that the inner belt contributes significantly in the mid-IR. Nonetheless, with the assumptions made for the modeling (same dust properties and density distribution for both belts), we do not see residual emission that could suggest the need for a third belt, as suggested by the SPHERE observations. We hypothesize that we can successfully model the SED of HIP\,73145 with the B1 and B3 belts.
Figure\,\ref{fig:sed_proba} shows the probability distributions for the results of the modeling. All parameters are well constrained and we derive uncertainties as explained in Section\,\ref{sec:disk_modeling}. Table\,\ref{tab:sed} summarizes our results. We find the disk to be slightly more compact that what we infer from the modeling of the SPHERE observations. The grain size distribution has an exponent in $-3.4$ that is close to the expected value for a collisional cascade ($-3.5$, \citealp{Dohnanyi1969}), and we find that the minimum grain size is of about $8.1$\,$\mu$m. The dust composition is dominated by carbon grains (80\%) with little to no water ices and a negligible porosity. We find that the inner dust belt should be located at a reference radius of about $\sim 5.5$\,au. The dust masses (between $s_{\mathrm{min}}$ and $s_{\mathrm{max}}$) are of $2.5 \times 10^{-2}$\,M$_{\oplus}$ and $7.2$\,M$_\oplus$ for the inner and outer belts, respectively. We computed the radially integrated optical depth for both the inner and outer belts. At the wavelength where the stellar emission peaks ($0.4$\,$\mu$m), we find optical depths of $\tau = 6.4 \times 10^{-3}$ and $1.4 \times 10^{-2}$ for the inner and outer belts, respectively. Therefore, according to our model, the debris disk is indeed optical thin in the radial direction, despite the significant dust mass we find for the outer belt.
This dust mass is significantly higher than the masses reported earlier by \citet{Hung2015b} of 0.27\,M$_\oplus$ and \citet{Moor2015} of 0.5\,M$_\oplus$. It clearly makes HIP73145 stand out even more peculiar against normal debris disks.

There is no requirement to introduce more complexity into the modeling in order to match the SED.  The B2 component, which does not appear to be very bright and probably does not contain too much dust anyway, was thus not introduced into the model. Degeneracy between the dust location and dust properties prohibits an attempt to match more detailed structures to the SED model, and the presence of B2 (and potentially more sub-structures) does not have a great impact on the overall SED. The same explanation holds with respect to the results of \citet{Hung2015} who fitted a three-component model, a continuous disk and two rings, to match the SED. The B3 and partially the B1 ring match the model well, as can be seen in Appendix~\ref{sec:sed-b1b3}.

\subsection{Discussion}

\begin{figure}
\includegraphics[width=1.0\columnwidth]{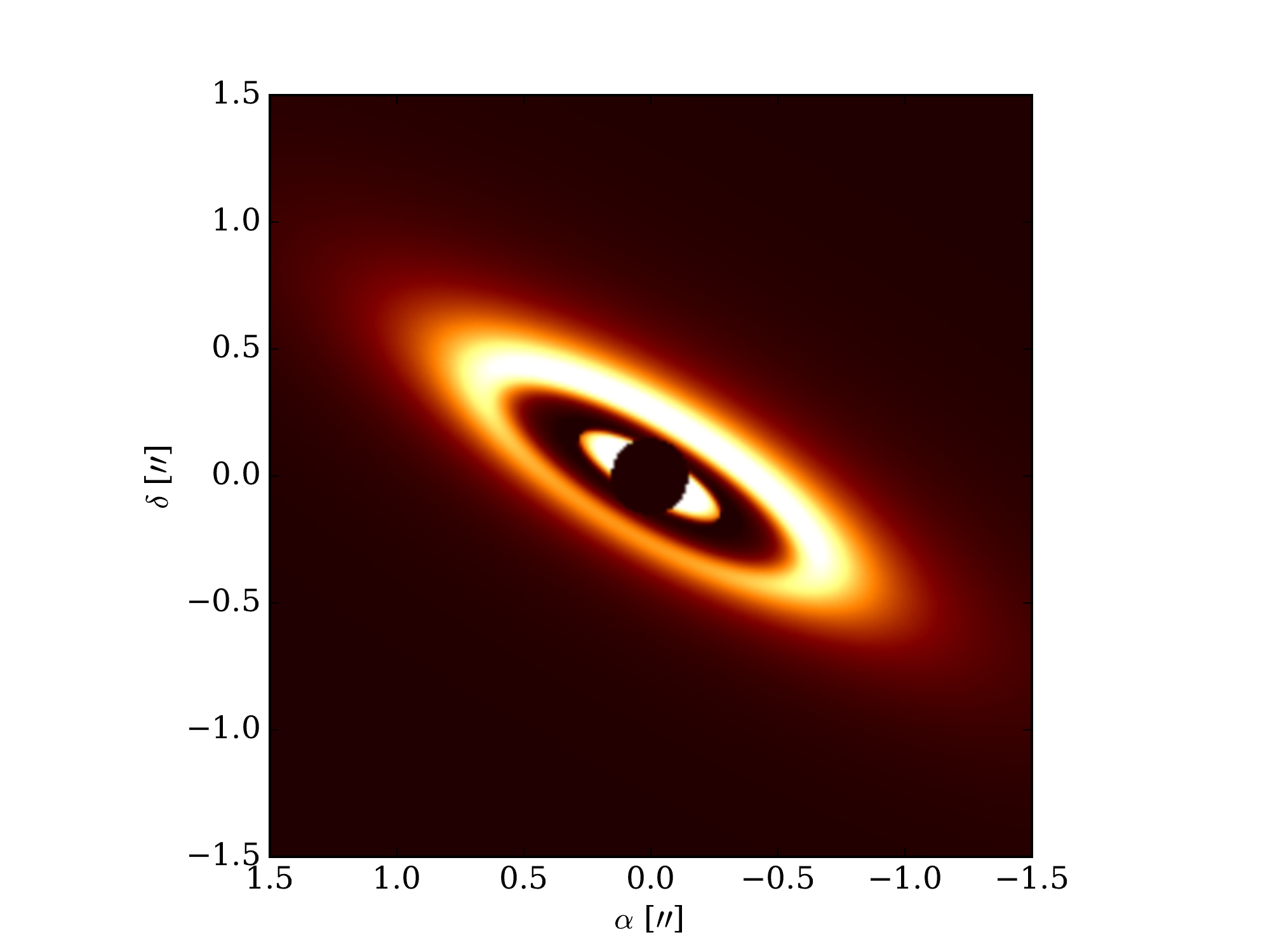}
\caption{Synthetic image including the inner and outer belt at the wavelength of the SPHERE observations with a central mask of radius $0^{\prime\prime}.15$.}
\label{fig:inner_disk}
\end{figure}

Even though this is highly speculative, we checked whether the inner belt, inferred from the SED modeling, could be detected with the SPHERE observations of the debris disk. We computed two images, one for the belt B1 (the best-fit model derived in Section\,\ref{sec:disk_modeling}) and another one for the B3 belt. The only differences between the two images are the reference radius ($5.4$\,au for B3) and the dust mass in the disk. All the other parameters ($i$, $\phi$, $\alpha_{\mathrm{in}}$, $\alpha_{\mathrm{out}}$, $\psi$, and $g$) are exactly the same. For the dust masses, we used the ratio derived from the modeling of the SED, which is a good first order approximation. We then use a central mask of radius $0^{\prime\prime}.15$ on the final image, which is shown in Figure\,\ref{fig:inner_disk}. Even though the reference radius for the inner belt would lie below the coronagraph used during the observations, the outermost regions of the inner disk still show up. This remains highly dependent on the outer slope of the density distribution and the dust mass ratio between the two disks, and would have to be compared to higher angular resolution observations. But the B1 belt detected with the SPHERE observations could be responsible for the emission at $20$\,$\mu$m. It should be noted that of course the SED modeling is highly degenerate, and while we obtain a good fit to the SED here, we can by no means actually constrain the ring structure by modeling the SED.  It also appears to make little sense to include the B2 ring in the SED model, which would not only add more free parameters and thus provide a better fit without actually adding constraints, but which also is expected to have a very similar dust temperature to B1.  From this perspective, the main motivation to include B3 is to include the "extreme" (in terms of dust temperature) rings B1 and B3, and to verify that the existence of B3 is not excluded by the SED model.

For the best-fit model dust composition, we compute the blowout size $s_{\mathrm{blow}}$ by radiation pressure as in \citet{Burns1979}. The size $s_{\mathrm{blow}}$ is defined as the size for which the ratio $\beta$ between the radiation pressure and gravitational forces equals $0.5$. Grains with larger $\beta$ values would be placed on unbound orbits, while grains with smaller ratios would be placed on bound orbits, provided their parent bodies are on circular orbits. We find that $s_{\mathrm{blow}}$ should be close to $8.6$\,$\mu$m (assuming a stellar mass of $1.77$\,M$_{\odot}$ and a luminosity of $11.5$\,L$_{\odot}$). The good agreement between this value and the $s_{\mathrm{min}}$ inferred during the modeling echoes the results of \citet[][the inferred value of $s_{\mathrm{min}}$ is close to $s_{\mathrm{blow}}$ for luminous stars]{Pawellek2014}.

The dust composition seems to be dominated by carbon. However this should not be overinterpreted, since there could be two explanations for this result. First, the absorption efficiencies of amorphous silicate grains of olivine stoichiometry are relatively low in the near-IR. Adding carbon increases the absorption efficiencies, and the grains become warmer, which could help reproduce the shape of the SED in the far-IR and sub-mm. Furthermore, the IRS spectrum does not show any emission features associated with silicate grains. Therefore, when including the inner belt, some of the grains could reach temperatures high enough to contribute at $10$\,$\mu$m. Carbon grains presenting featureless absorption efficiencies consequently help reproduce the shape of the IRS spectrum.

\section{Conclusions}

  We detect a structure of concentric rings around HIP\,73145. To verify that the closer in ring-like structures are not artifacts from the ADI data processing, we injected single-component models into the data and ran them through the analysis pipeline. No ring-like structures whatsoever were created by the data analysis process from single component models.

  Such structures are unusual in debris disks, however the gas mass of (much) more than 0.06\,M$_{\oplus}$ estimates from \citet{Moor2016} and the dust mass of more than 7.2\,M$_{\oplus}$ determined in this work point to an object, which retained a large amount of gas and dust despite of its well-determined age of 15\,Myr.  HIP\,73145 could thus be a member of a somewhat intermediate class of objects that retain a relatively high gas and dust mass at relatively old ages.

A deep companion search yields a faint candidate outside the disk, about 2\farcs2 from HIP\,73145, most likely
a background object, and allows us to exclude the presence of planetary companions more massive than about 2.5\,M$_\mathrm{Jup}$ outside of 0\farcs2 or 25\,AU.

  Modeling the SPHERE and ALMA observations, we determine the dust distribution of two different components.  When assuming that the reference radius in both models should be the same at 90\,AU, it is clear that the distribution of smaller dust grains detected by SPHERE extends further outward than that of the larger grains traced by ALMA.

  The SED fitting requires multiple components, which makes HIP\,73145 one of the rare objects where actually detected multiple scattering components are present and required by the SED model fit.  Still it is not fully determined how the detected B3 component is actually related to the required hot inner belt of the SED model.  The B2 component detected in scattered light is not very bright and does not have sufficient impact on the SED model to use the latter as independent confirmation of its presence.

  Whether the substructures in the HIP\,73145 disk are caused by dynamical interactions with planets or by dust-gas interactions cannot be clarified in this paper.

\begin{acknowledgements}

This paper is based on observations collected at the European Organisation for Astronomical Research in the southern hemisphere under ESO programmme 095.C-0298.

SPHERE is an instrument designed and built by a consortium consisting of IPAG (Grenoble, France), MPIA (Heidelberg, Germany), LAM (Marseille, France), LESIA (Paris, France), Laboratoire Lagrange (Nice, France), INAF - Osservatorio di Padova (Italy), Observatoire de Gen-
eve (Switzerland), ETH Zurich (Switzerland), NOVA (Netherlands), ONERA
(France), and ASTRON (Netherlands) in collaboration with ESO. SPHERE was
funded by ESO, with additional contributions from CNRS (France), MPIA (Germany), INAF (Italy), FINES (Switzerland), and NOVA (Netherlands). SPHERE
has also received funding from the European Commission Sixth and Seventh Framework Programmes as part of the Optical Infrared Coordination Network for Astronomy (OPTICON) under grant number RII3-Ct-2004-001566 for FP6 (2004-2008), grant number 226604 for FP7 (2009-2012) and grant number 312430 for
FP7 (2013-2016).

This paper makes use of the following ALMA data: ADS/JAO.ALMA 2012.1.00688.S. ALMA is a partnership of ESO (representing its member states), NSF (USA), and NINS (Japan) together with NRC (Canada), NSC and ASIAA (Taiwan), and KASI (Republic of Korea), in cooperation with the Republic of Chile. The Joint ALMA Observatory is operated by ESO, AUI/NRAO, and NAOJ. J. Olofsson acknowledges support from the Millennium Nucleus RC130007 (Chilean Ministry of Economy).

S. Desidera, R. Gratton, D. Mesa, and E. Sissa acknowledge support from the “Progetti Premiali” funding scheme of the Italian Ministry of Education, University, and Research.

J.O acknowledges support from ALMA/Conicyt Project 31130027, and from the Millennium Nucleus RC130007 (Chilean Ministry of Economy).

A. Moor's work was supported by the Momentum grant of the MTA CSFK Lend\"ulet Disk Research Group.

We thank the anonymous referee for providing helpful comments and contributing greatly to the improvement of the paper.
\end{acknowledgements}

\bibliographystyle{aa}
\bibliography{biblio}

\begin{thebibliography}{64}
\expandafter\ifx\csname natexlab\endcsname\relax\def\natexlab#1{#1}\fi

\bibitem[{Albert {et~al.}(2008)Albert, Meunier, \& Delorme}]{SPHERE-DC}
Albert, D., Meunier, N., \& Delorme, P. 2008, SPHERE Data Centre,
  \url{http://sphere.osug.fr/spip.php?rubrique16&lang=en}, accessed: 2016-06-09

\bibitem[{{Allers} \& {Liu}(2013)}]{2013ApJ...772...79A}
{Allers}, K.~N. \& {Liu}, M.~C. 2013, \apj, 772, 79

\bibitem[{{Baraffe} {et~al.}(2003){Baraffe}, {Chabrier}, {Barman}, {Allard}, \&
  {Hauschildt}}]{Baraffe2003}
{Baraffe}, I., {Chabrier}, G., {Barman}, T.~S., {Allard}, F., \& {Hauschildt},
  P.~H. 2003, \aap, 402, 701

\bibitem[{{Baraffe} {et~al.}(2015){Baraffe}, {Homeier}, {Allard}, \&
  {Chabrier}}]{Baraffe2015}
{Baraffe}, I., {Homeier}, D., {Allard}, F., \& {Chabrier}, G. 2015, \aap, 577,
  A42

\bibitem[{{Beuzit} {et~al.}(2008){Beuzit}, {Feldt}, {Dohlen}, {Mouillet},
  {Puget}, {Wildi}, {Abe}, {Antichi}, {Baruffolo}, {Baudoz}, {Boccaletti},
  {Carbillet}, {Charton}, {Claudi}, {Downing}, {Fabron}, {Feautrier},
  {Fedrigo}, {Fusco}, {Gach}, {Gratton}, {Henning}, {Hubin}, {Joos}, {Kasper},
  {Langlois}, {Lenzen}, {Moutou}, {Pavlov}, {Petit}, {Pragt}, {Rabou}, {Rigal},
  {Roelfsema}, {Rousset}, {Saisse}, {Schmid}, {Stadler}, {Thalmann}, {Turatto},
  {Udry}, {Vakili}, \& {Waters}}]{Beuzit2008}
{Beuzit}, J.-L., {Feldt}, M., {Dohlen}, K., {et~al.} 2008, in \procspie, Vol.
  7014, Ground-based and Airborne Instrumentation for Astronomy II, 701418

\bibitem[{{Boccaletti} {et~al.}(2008){Boccaletti}, {Abe}, {Baudrand}, {Daban},
  {Douet}, {Guerri}, {Robbe-Dubois}, {Bendjoya}, {Dohlen}, \&
  {Mawet}}]{Boccaletti2008}
{Boccaletti}, A., {Abe}, L., {Baudrand}, J., {et~al.} 2008, in \procspie, Vol.
  7015, Adaptive Optics Systems, 70151B

\bibitem[{{Boccaletti} {et~al.}(2015){Boccaletti}, {Thalmann}, {Lagrange},
  {Janson}, {Augereau}, {Schneider}, {Milli}, {Grady}, {Debes}, {Langlois},
  {Mouillet}, {Henning}, {Dominik}, {Maire}, {Beuzit}, {Carson}, {Dohlen},
  {Engler}, {Feldt}, {Fusco}, {Ginski}, {Girard}, {Hines}, {Kasper}, {Mawet},
  {M{\'e}nard}, {Meyer}, {Moutou}, {Olofsson}, {Rodigas}, {Sauvage},
  {Schlieder}, {Schmid}, {Turatto}, {Udry}, {Vakili}, {Vigan}, {Wahhaj}, \&
  {Wisniewski}}]{Boccaletti2015}
{Boccaletti}, A., {Thalmann}, C., {Lagrange}, A.-M., {et~al.} 2015, \nat, 526,
  230

\bibitem[{{Bonnefoy} {et~al.}(2014){Bonnefoy}, {Chauvin}, {Lagrange}, {Rojo},
  {Allard}, {Pinte}, {Dumas}, \& {Homeier}}]{2014A&A...562A.127B}
{Bonnefoy}, M., {Chauvin}, G., {Lagrange}, A.-M., {et~al.} 2014, \aap, 562,
  A127

\bibitem[{{Burgasser}(2014)}]{2014ASInC..11....7B}
{Burgasser}, A.~J. 2014, in Astronomical Society of India Conference Series,
  Vol.~11, Astronomical Society of India Conference Series

\bibitem[{{Burgasser} {et~al.}(2010){Burgasser}, {Simcoe}, {Bochanski},
  {Saumon}, {Mamajek}, {Cushing}, {Marley}, {McMurtry}, {Pipher}, \&
  {Forrest}}]{2010ApJ...725.1405B}
{Burgasser}, A.~J., {Simcoe}, R.~A., {Bochanski}, J.~J., {et~al.} 2010, \apj,
  725, 1405

\bibitem[{{Burns} {et~al.}(1979){Burns}, {Lamy}, \& {Soter}}]{Burns1979}
{Burns}, J.~A., {Lamy}, P.~L., \& {Soter}, S. 1979, \icarus, 40, 1

\bibitem[{{Castelli} {et~al.}(1997){Castelli}, {Gratton}, \&
  {Kurucz}}]{Castelli1997}
{Castelli}, F., {Gratton}, R.~G., \& {Kurucz}, R.~L. 1997, \aap, 318, 841

\bibitem[{{Chen} {et~al.}(2014){Chen}, {Mittal}, {Kuchner}, {Forrest}, {Lisse},
  {Manoj}, {Sargent}, \& {Watson}}]{Chen2014}
{Chen}, C.~H., {Mittal}, T., {Kuchner}, M., {et~al.} 2014, \apjs, 211, 25

\bibitem[{{Chen} {et~al.}(2012){Chen}, {Pecaut}, {Mamajek}, {Su}, \&
  {Bitner}}]{Chen2012}
{Chen}, C.~H., {Pecaut}, M., {Mamajek}, E.~E., {Su}, K.~Y.~L., \& {Bitner}, M.
  2012, \apj, 756, 133

\bibitem[{{Claudi} {et~al.}(2008){Claudi}, {Turatto}, {Gratton}, {Antichi},
  {Bonavita}, {Bruno}, {Cascone}, {De Caprio}, {Desidera}, {Giro}, {Mesa},
  {Scuderi}, {Dohlen}, {Beuzit}, \& {Puget}}]{Claudi2008}
{Claudi}, R.~U., {Turatto}, M., {Gratton}, R.~G., {et~al.} 2008, in \procspie,
  Vol. 7014, Ground-based and Airborne Instrumentation for Astronomy II, 70143E

\bibitem[{{Delorme} {et~al.}(2008){Delorme}, {Delfosse}, {Albert}, {Artigau},
  {Forveille}, {Reyl{\'e}}, {Allard}, {Homeier}, {Robin}, {Willott}, {Liu}, \&
  {Dupuy}}]{2008A&A...482..961D}
{Delorme}, P., {Delfosse}, X., {Albert}, L., {et~al.} 2008, \aap, 482, 961

\bibitem[{{Dipierro} {et~al.}(2015){Dipierro}, {Price}, {Laibe}, {Hirsh},
  {Cerioli}, \& {Lodato}}]{Dipierro2015}
{Dipierro}, G., {Price}, D., {Laibe}, G., {et~al.} 2015, \mnras, 453, L73

\bibitem[{{Dohnanyi}(1969)}]{Dohnanyi1969}
{Dohnanyi}, J.~S. 1969, \jgr, 74, 2531

\bibitem[{{Dorschner} {et~al.}(1995){Dorschner}, {Begemann}, {Henning},
  {J{\"a}ger}, \& {Mutschke}}]{Dorschner1995}
{Dorschner}, J., {Begemann}, B., {Henning}, T., {J{\"a}ger}, C., \& {Mutschke},
  H. 1995, \aap, 300, 503

\bibitem[{{Flock} {et~al.}(2015){Flock}, {Ruge}, {Dzyurkevich}, {Henning},
  {Klahr}, \& {Wolf}}]{Flock2015}
{Flock}, M., {Ruge}, J.~P., {Dzyurkevich}, N., {et~al.} 2015, \aap, 574, A68

\bibitem[{{Foreman-Mackey} {et~al.}(2012){Foreman-Mackey}, {Hogg}, {Lang}, \&
  {Goodman}}]{Foreman-Mackey2012}
{Foreman-Mackey}, D., {Hogg}, D.~W., {Lang}, D., \& {Goodman}, J. 2012, ArXiv
  e-prints, 1202.3665

\bibitem[{{Fusco} {et~al.}(2014){Fusco}, {Sauvage}, {Petit}, {Costille},
  {Dohlen}, {Mouillet}, {Beuzit}, {Kasper}, {Suarez}, {Soenke}, {Fedrigo},
  {Downing}, {Baudoz}, {Sevin}, {Perret}, {Barrufolo}, {Salasnich}, {Puget},
  {Feautrier}, {Rochat}, {Moulin}, {Deboulb{\'e}}, {Hugot}, {Vigan}, {Mawet},
  {Girard}, \& {Hubin}}]{Fusco2014}
{Fusco}, T., {Sauvage}, J.-F., {Petit}, C., {et~al.} 2014, in \procspie, Vol.
  9148, Adaptive Optics Systems IV, 91481U

\bibitem[{{Gauza} {et~al.}(2015){Gauza}, {B{\'e}jar}, {P{\'e}rez-Garrido},
  {Rosa Zapatero Osorio}, {Lodieu}, {Rebolo}, {Pall{\'e}}, \&
  {Nowak}}]{2015ApJ...804...96G}
{Gauza}, B., {B{\'e}jar}, V.~J.~S., {P{\'e}rez-Garrido}, A., {et~al.} 2015,
  \apj, 804, 96

\bibitem[{{Gizis} {et~al.}(2015){Gizis}, {Allers}, {Liu}, {Harris}, {Faherty},
  {Burgasser}, \& {Kirkpatrick}}]{2015ApJ...799..203G}
{Gizis}, J.~E., {Allers}, K.~N., {Liu}, M.~C., {et~al.} 2015, \apj, 799, 203

\bibitem[{{Gonzalez} {et~al.}(2015){Gonzalez}, {Laibe}, {Maddison}, {Pinte}, \&
  {M{\'e}nard}}]{Gonzalez2015}
{Gonzalez}, J.-F., {Laibe}, G., {Maddison}, S.~T., {Pinte}, C., \&
  {M{\'e}nard}, F. 2015, \mnras, 454, L36

\bibitem[{{Henyey} \& {Greenstein}(1941)}]{Henyey1941}
{Henyey}, L.~G. \& {Greenstein}, J.~L. 1941, \apj, 93, 70

\bibitem[{{Houk}(1982)}]{Houk1982}
{Houk}, N. 1982, {Michigan Catalogue of Two-dimensional Spectral Types for the
  HD stars. Volume\_3. Declinations -40\degr to -26\degr}

\bibitem[{Hung {et~al.}(2015b)Hung, Duchêne, Arriaga, Fitzgerald, Maire,
  Marois, Millar-Blanchaer, Bruzzone, Rajan, Pueyo, Kalas, Rosa, Graham,
  Konopacky, Wolff, Ammons, Chen, Chilcote, Draper, Esposito, Gerard, Goodsell,
  Greenbaum, Hibon, Hinkley, Macintosh, Marchis, Metchev, Nielsen, Oppenheimer,
  Patience, Perrin, Rantakyrö, Sivaramakrishnan, Wang, Ward-Duong, \&
  Wiktorowicz}]{Hung2015b}
Hung, L.-W., Duchêne, G., Arriaga, P., {et~al.} 2015b, The Astrophysical
  Journal Letters, 815, L14

\bibitem[{{Hung} {et~al.}(2015a){Hung}, {Fitzgerald}, {Chen}, {Mittal},
  {Kalas}, \& {Graham}}]{Hung2015}
{Hung}, L.-W., {Fitzgerald}, M.~P., {Chen}, C.~H., {et~al.} 2015a, \apj, 802,
  138

\bibitem[{{J{\"a}ger} {et~al.}(1998){J{\"a}ger}, {Mutschke}, \&
  {Henning}}]{Jager1998}
{J{\"a}ger}, C., {Mutschke}, H., \& {Henning}, T. 1998, \aap, 332, 291

\bibitem[{Klahr \& Lin(2005)}]{Klahr2005}
Klahr, H. \& Lin, D. N.~C. 2005, The Astrophysical Journal, 632, 1113

\bibitem[{{Lafreni{\`e}re} {et~al.}(2010){Lafreni{\`e}re}, {Jayawardhana}, \&
  {van Kerkwijk}}]{2010ApJ...719..497L}
{Lafreni{\`e}re}, D., {Jayawardhana}, R., \& {van Kerkwijk}, M.~H. 2010, \apj,
  719, 497

\bibitem[{{Lafreni{\`e}re} {et~al.}(2007){Lafreni{\`e}re}, {Marois}, {Doyon},
  {Nadeau}, \& {Artigau}}]{Lafreniere2007}
{Lafreni{\`e}re}, D., {Marois}, C., {Doyon}, R., {Nadeau}, D., \& {Artigau},
  {\'E}. 2007, \apj, 660, 770

\bibitem[{{Li} \& {Greenberg}(1998)}]{Li1998}
{Li}, A. \& {Greenberg}, J.~M. 1998, \aap, 331, 291

\bibitem[{{Liu} {et~al.}(2013){Liu}, {Magnier}, {Deacon}, {Allers}, {Dupuy},
  {Kotson}, {Aller}, {Burgett}, {Chambers}, {Draper}, {Hodapp}, {Jedicke},
  {Kaiser}, {Kudritzki}, {Metcalfe}, {Morgan}, {Price}, {Tonry}, \&
  {Wainscoat}}]{2013ApJ...777L..20L}
{Liu}, M.~C., {Magnier}, E.~A., {Deacon}, N.~R., {et~al.} 2013, \apjl, 777, L20

\bibitem[{{Lyra} \& {Kuchner}(2013)}]{Lyra2013}
{Lyra}, W. \& {Kuchner}, M. 2013, \nat, 499, 184

\bibitem[{{Mace} {et~al.}(2013){Mace}, {Kirkpatrick}, {Cushing}, {Gelino},
  {Griffith}, {Skrutskie}, {Marsh}, {Wright}, {Eisenhardt}, {McLean},
  {Thompson}, {Mix}, {Bailey}, {Beichman}, {Bloom}, {Burgasser}, {Fortney},
  {Hinz}, {Knox}, {Lowrance}, {Marley}, {Morley}, {Rodigas}, {Saumon},
  {Sheppard}, \& {Stock}}]{2013ApJS..205....6M}
{Mace}, G.~N., {Kirkpatrick}, J.~D., {Cushing}, M.~C., {et~al.} 2013, \apjs,
  205, 6

\bibitem[{{Macintosh} {et~al.}(2015){Macintosh}, {Graham}, {Barman}, {De Rosa},
  {Konopacky}, {Marley}, {Marois}, {Nielsen}, {Pueyo}, {Rajan}, {Rameau},
  {Saumon}, {Wang}, {Patience}, {Ammons}, {Arriaga}, {Artigau}, {Beckwith},
  {Brewster}, {Bruzzone}, {Bulger}, {Burningham}, {Burrows}, {Chen}, {Chiang},
  {Chilcote}, {Dawson}, {Dong}, {Doyon}, {Draper}, {Duch{\^e}ne}, {Esposito},
  {Fabrycky}, {Fitzgerald}, {Follette}, {Fortney}, {Gerard}, {Goodsell},
  {Greenbaum}, {Hibon}, {Hinkley}, {Cotten}, {Hung}, {Ingraham},
  {Johnson-Groh}, {Kalas}, {Lafreniere}, {Larkin}, {Lee}, {Line}, {Long},
  {Maire}, {Marchis}, {Matthews}, {Max}, {Metchev}, {Millar-Blanchaer},
  {Mittal}, {Morley}, {Morzinski}, {Murray-Clay}, {Oppenheimer}, {Palmer},
  {Patel}, {Perrin}, {Poyneer}, {Rafikov}, {Rantakyr{\"o}}, {Rice}, {Rojo},
  {Rudy}, {Ruffio}, {Ruiz}, {Sadakuni}, {Saddlemyer}, {Salama}, {Savransky},
  {Schneider}, {Sivaramakrishnan}, {Song}, {Soummer}, {Thomas}, {Vasisht},
  {Wallace}, {Ward-Duong}, {Wiktorowicz}, {Wolff}, \&
  {Zuckerman}}]{2015Sci...350...64M}
{Macintosh}, B., {Graham}, J.~R., {Barman}, T., {et~al.} 2015, Science, 350, 64

\bibitem[{{Maire} {et~al.}(2016){Maire}, {Bonnefoy}, {Ginski}, {Vigan},
  {Messina}, {Mesa}, {Galicher}, {Gratton}, {Desidera}, {Kopytova}, {Millward},
  {Thalmann}, {Claudi}, {Ehrenreich}, {Zurlo}, {Chauvin}, {Antichi},
  {Baruffolo}, {Bazzon}, {Beuzit}, {Blanchard}, {Boccaletti}, {de Boer},
  {Carle}, {Cascone}, {Costille}, {De Caprio}, {Delboulb{\'e}}, {Dohlen},
  {Dominik}, {Feldt}, {Fusco}, {Girard}, {Giro}, {Gisler}, {Gluck}, {Gry},
  {Henning}, {Hubin}, {Hugot}, {Jaquet}, {Kasper}, {Lagrange}, {Langlois}, {Le
  Mignant}, {Llored}, {Madec}, {Martinez}, {Mawet}, {Milli},
  {M{\"o}ller-Nilsson}, {Mouillet}, {Moulin}, {Moutou}, {Orign{\'e}}, {Pavlov},
  {Petit}, {Pragt}, {Puget}, {Ramos}, {Rochat}, {Roelfsema}, {Salasnich},
  {Sauvage}, {Schmid}, {Turatto}, {Udry}, {Vakili}, {Wahhaj}, {Weber}, \&
  {Wildi}}]{Maire2016}
{Maire}, A.-L., {Bonnefoy}, M., {Ginski}, C., {et~al.} 2016, \aap, 587, A56

\bibitem[{{Marois} {et~al.}(2014){Marois}, {Correia}, {Galicher}, {Ingraham},
  {Macintosh}, {Currie}, \& {De Rosa}}]{Marois2014}
{Marois}, C., {Correia}, C., {Galicher}, R., {et~al.} 2014, in \procspie, Vol.
  9148, Adaptive Optics Systems IV, 91480U

\bibitem[{{Marois} {et~al.}(2006){Marois}, {Lafreni{\`e}re}, {Doyon},
  {Macintosh}, \& {Nadeau}}]{Marois2006}
{Marois}, C., {Lafreni{\`e}re}, D., {Doyon}, R., {Macintosh}, B., \& {Nadeau},
  D. 2006, \apj, 641, 556

\bibitem[{{McMullin} {et~al.}(2007){McMullin}, {Waters}, {Schiebel}, {Young},
  \& {Golap}}]{McMullin2007}
{McMullin}, J.~P., {Waters}, B., {Schiebel}, D., {Young}, W., \& {Golap}, K.
  2007, in Astronomical Society of the Pacific Conference Series, Vol. 376,
  Astronomical Data Analysis Software and Systems XVI, ed. R.~A. {Shaw},
  F.~{Hill}, \& D.~J. {Bell}, 127

\bibitem[{Milli {et~al.}(2012)Milli, Mouillet, Lagrange, Boccaletti, Mawet,
  Chauvin, \& Bonnefoy}]{Milli2012}
Milli, J., Mouillet, D., Lagrange, A.~M., {et~al.} 2012, Astronomy {\&}
  Astrophysics, 545, A111

\bibitem[{Moor(2016)}]{Moor2016}
Moor, A. 2016, in preparation

\bibitem[{{Mo{\'o}r} {et~al.}(2006){Mo{\'o}r}, {{\'A}brah{\'a}m}, {Derekas},
  {Kiss}, {Kiss}, {Apai}, {Grady}, \& {Henning}}]{Moor2006}
{Mo{\'o}r}, A., {{\'A}brah{\'a}m}, P., {Derekas}, A., {et~al.} 2006, \apj, 644,
  525

\bibitem[{{Mo{\'o}r} {et~al.}(2015){Mo{\'o}r}, {Henning}, {Juh{\'a}sz},
  {{\'A}brah{\'a}m}, {Balog}, {K{\'o}sp{\'a}l}, {Pascucci}, {Szab{\'o}},
  {Vavrek}, {Cur{\'e}}, {Csengeri}, {Grady}, {G{\"u}sten}, \&
  {Kiss}}]{Moor2015}
{Mo{\'o}r}, A., {Henning}, T., {Juh{\'a}sz}, A., {et~al.} 2015, \apj, 814, 42

\bibitem[{{Okuzumi} {et~al.}(2015){Okuzumi}, {Momose}, {Sirono}, {Kobayashi},
  \& {Tanaka}}]{Okuzumi2016}
{Okuzumi}, S., {Momose}, M., {Sirono}, S.-i., {Kobayashi}, H., \& {Tanaka}, H.
  2015, ArXiv e-prints, 1510.03556

\bibitem[{{Olofsson} {et~al.}(2016){Olofsson}, {Samland}, {Avenhaus},
  {Caceres}, {Henning}, {Moor}, {Milli}, {Canovas}, {Quanz}, {Schreiber},
  {Augereau}, {Bayo}, {Bazzon}, {Beuzit}, {Boccaletti}, {Buenzli}, {Casassus},
  {Chauvin}, {Dominik}, {Desidera}, {Feldt}, {Gratton}, {Janson}, {Lagrange},
  {Langlois}, {Lannier}, {Maire}, {Mesa}, {Pinte}, {Rouan}, {Salter},
  {Thalmann}, \& {Vigan}}]{Olofsson2016}
{Olofsson}, J., {Samland}, M., {Avenhaus}, H., {et~al.} 2016, ArXiv e-prints,
  1601.07861

\bibitem[{{Patience} {et~al.}(2010){Patience}, {King}, {de Rosa}, \&
  {Marois}}]{2010A&A...517A..76P}
{Patience}, J., {King}, R.~R., {de Rosa}, R.~J., \& {Marois}, C. 2010, \aap,
  517, A76

\bibitem[{{Pavlov} {et~al.}(2008){Pavlov}, {M{\"o}ller-Nilsson}, {Feldt},
  {Henning}, {Beuzit}, \& {Mouillet}}]{Pavlov2008}
{Pavlov}, A., {M{\"o}ller-Nilsson}, O., {Feldt}, M., {et~al.} 2008, in
  \procspie, Vol. 7019, Advanced Software and Control for Astronomy II, 701939

\bibitem[{Pawellek {et~al.}(2014)Pawellek, Krivov, Marshall, Montesinos,
  Abraham, Moor, Bryden, \& Eiroa}]{Pawellek2014}
Pawellek, N., Krivov, A.~V., Marshall, J.~P., {et~al.} 2014, The Astrophysical
  Journal, 792, 65

\bibitem[{{Pecaut} {et~al.}(2012){Pecaut}, {Mamajek}, \& {Bubar}}]{Pecaut2012}
{Pecaut}, M.~J., {Mamajek}, E.~E., \& {Bubar}, E.~J. 2012, \apj, 746, 154

\bibitem[{Pedregosa {et~al.}(2011)Pedregosa, Varoquaux, Gramfort, Michel,
  Thirion, Grisel, Blondel, Prettenhofer, Weiss, Dubourg, Vanderplas, Passos,
  Cournapeau, Brucher, Perrot, \& Duchesnay}]{Pedregosa2011}
Pedregosa, F., Varoquaux, G., Gramfort, A., {et~al.} 2011, Journal of Machine
  Learning Research, 12, 2825

\bibitem[{{Picogna} \& {Kley}(2015)}]{Picogna2015}
{Picogna}, G. \& {Kley}, W. 2015, \aap, 584, A110

\bibitem[{{Rizzuto} {et~al.}(2011){Rizzuto}, {Ireland}, \&
  {Robertson}}]{Rizzuto2011}
{Rizzuto}, A.~C., {Ireland}, M.~J., \& {Robertson}, J.~G. 2011, \mnras, 416,
  3108

\bibitem[{{Schneider} {et~al.}(2015){Schneider}, {Cushing}, {Kirkpatrick},
  {Gelino}, {Mace}, {Wright}, {Eisenhardt}, {Skrutskie}, {Griffith}, \&
  {Marsh}}]{2015ApJ...804...92S}
{Schneider}, A.~C., {Cushing}, M.~C., {Kirkpatrick}, J.~D., {et~al.} 2015,
  \apj, 804, 92

\bibitem[{{Schneider} {et~al.}(2014){Schneider}, {Cushing}, {Kirkpatrick},
  {Mace}, {Gelino}, {Faherty}, {Fajardo-Acosta}, \&
  {Sheppard}}]{2014AJ....147...34S}
{Schneider}, A.~C., {Cushing}, M.~C., {Kirkpatrick}, J.~D., {et~al.} 2014, \aj,
  147, 34

\bibitem[{{Soummer} {et~al.}(2012){Soummer}, {Pueyo}, \&
  {Larkin}}]{Soummer2012}
{Soummer}, R., {Pueyo}, L., \& {Larkin}, J. 2012, \apjl, 755, L28

\bibitem[{{Thalmann} {et~al.}(2011){Thalmann}, {Janson}, {Buenzli}, {Brandt},
  {Wisniewski}, {Moro-Mart{\'{\i}}n}, {Usuda}, {Schneider}, {Carson},
  {McElwain}, {Grady}, {Goto}, {Abe}, {Brandner}, {Dominik}, {Egner}, {Feldt},
  {Fukue}, {Golota}, {Guyon}, {Hashimoto}, {Hayano}, {Hayashi}, {Hayashi},
  {Henning}, {Hodapp}, {Ishii}, {Iye}, {Kandori}, {Knapp}, {Kudo}, {Kusakabe},
  {Kuzuhara}, {Matsuo}, {Miyama}, {Morino}, {Nishimura}, {Pyo}, {Serabyn},
  {Suto}, {Suzuki}, {Takahashi}, {Takami}, {Takato}, {Terada}, {Tomono},
  {Turner}, {Watanabe}, {Yamada}, {Takami}, \& {Tamura}}]{Thalmann2011}
{Thalmann}, C., {Janson}, M., {Buenzli}, E., {et~al.} 2011, \apjl, 743, L6

\bibitem[{{van Leeuwen}(2007)}]{vanLeeuwen2007}
{van Leeuwen}, F. 2007, \aap, 474, 653

\bibitem[{{Vigan} {et~al.}(2010){Vigan}, {Moutou}, {Langlois}, {Allard},
  {Boccaletti}, {Carbillet}, {Mouillet}, \& {Smith}}]{Vigan2010}
{Vigan}, A., {Moutou}, C., {Langlois}, M., {et~al.} 2010, \mnras, 407, 71

\bibitem[{{Wahhaj} {et~al.}(2011){Wahhaj}, {Liu}, {Biller}, {Clarke},
  {Nielsen}, {Close}, {Hayward}, {Mamajek}, {Cushing}, {Dupuy}, {Tecza},
  {Thatte}, {Chun}, {Ftaclas}, {Hartung}, {Reid}, {Shkolnik}, {Alencar},
  {Artymowicz}, {Boss}, {de Gouveia Dal Pino}, {Gregorio-Hetem}, {Ida},
  {Kuchner}, {Lin}, \& {Toomey}}]{2011ApJ...729..139W}
{Wahhaj}, Z., {Liu}, M.~C., {Biller}, B.~A., {et~al.} 2011, \apj, 729, 139

\bibitem[{{Wright} {et~al.}(2003){Wright}, {Egan}, {Kraemer}, \&
  {Price}}]{Wright2003}
{Wright}, C.~O., {Egan}, M.~P., {Kraemer}, K.~E., \& {Price}, S.~D. 2003, \aj,
  125, 359

\bibitem[{{Zhang} {et~al.}(2015){Zhang}, {Blake}, \& {Bergin}}]{Zhang2015}
{Zhang}, K., {Blake}, G.~A., \& {Bergin}, E.~A. 2015, \apjl, 806, L7

\end{thebibliography}

\begin{appendix}
  \section{Comparing modeling results to data}

  \subsection{The effect of data reduction on the B2 and B3 structures \label{sec:dr-b2b3}}

  \begin{figure}
  \includegraphics[width=1.0\columnwidth]{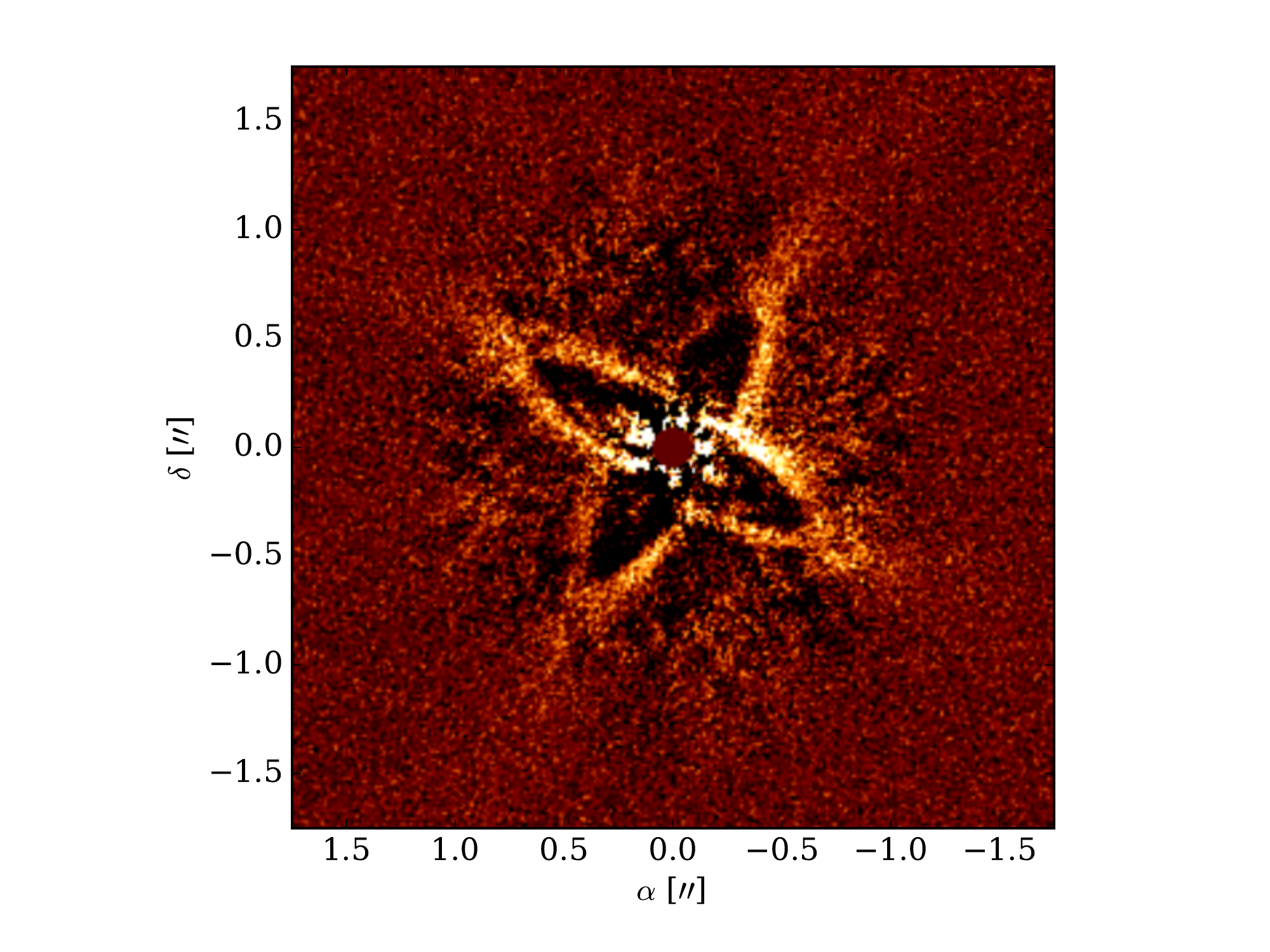}
  \caption{Best-fit model injected into the observations at a $90^{\circ}$ angle. The image is the average of image processed through PCA with 3, 5, 7, 10, and 15 components.}
  \label{fig:fake_disks}
  \end{figure}

  In Sec.~\ref{sec:disk-structure}, we discussed the presence of additional rings close to the star (B2 and B3). If confirmed, this would make HIP\,73145 one of the few debris disks with spatially resolved separated dust belts. Nonetheless, the ADI process can be quite aggressive on the final flux distribution, especially in the innermost regions, and we need to rule out the PCA process as the explanation for the bright spots close to the center of the image.

  To assess the effect of the PCA process on the innermost regions, we took the best-fit model and injected it in the observed data cube (the original, rather than the downsampled and resized cube), at the proper parallactic angles. Instead of subtracting the disk model, we added it, with an additional $90^{\circ}$ rotation (e.g., \citealp{Thalmann2011}). We used a numerical central mask with a radius of $0.0925^{\prime\prime}$ (comparable to the coronagraph used during the observations). We then ran the PCA process, keeping the 3, 5, 7, 10, and 15 principal components, and we averaged the results. The final image is shown in Figure\,\ref{fig:fake_disks}. Both the ``real'' and ``fake'' disks appear very similar. Nonetheless, we do not observe the same behavior close to the central mask; there are no counterparts to the B2 or B3 structures with a $90^{\circ}$ angle. We can therefore conclude that these structures are not the consequences of the PCA process and, therefore, may well be real.

  \subsection{The SED model components in scattered light \label{sec:sed-b1b3}}

  \begin{figure*}
  \includegraphics[width=2.0\columnwidth]{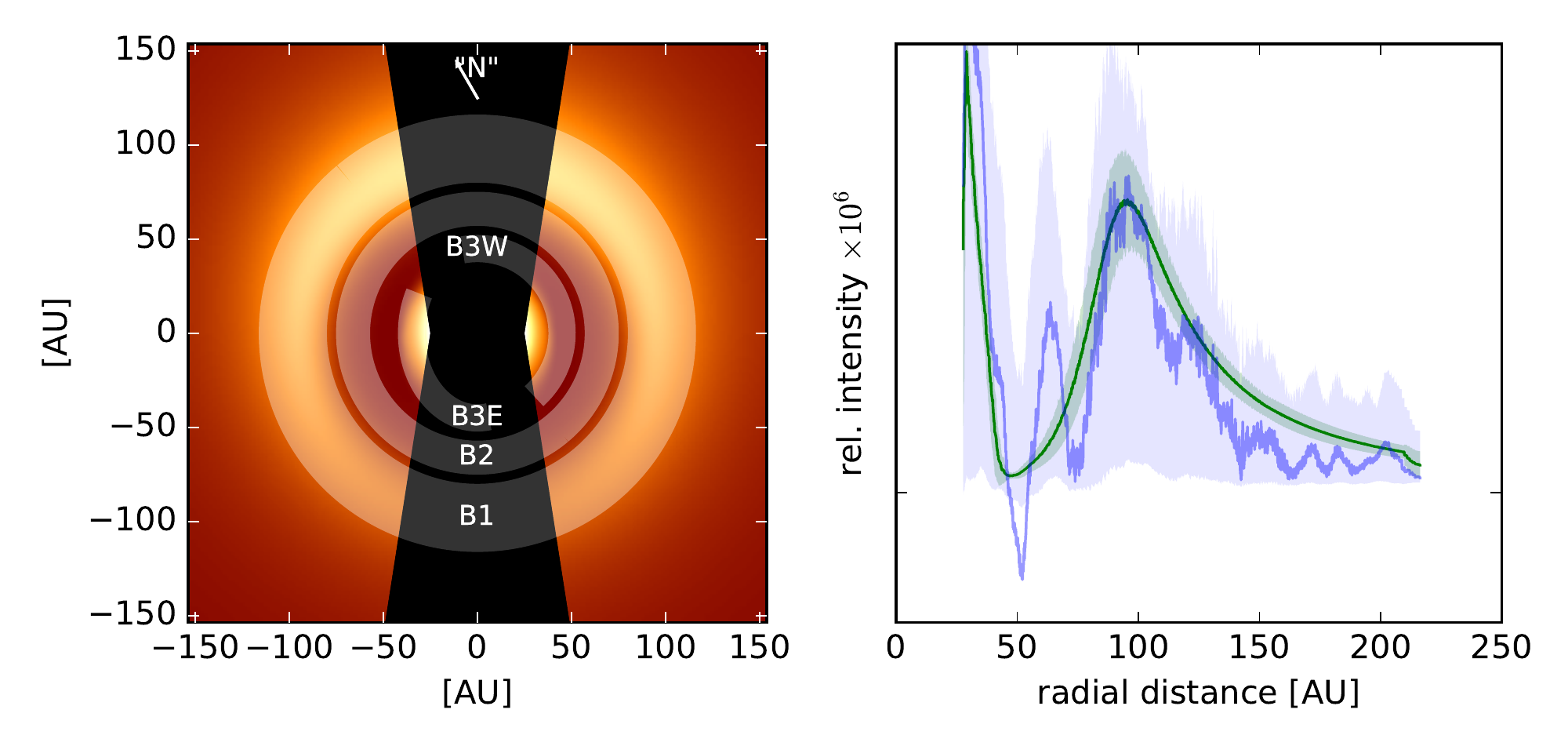}
  \caption{Same as Fig.~\ref{fig:deproj}, but derived from the model described in Sec.~\ref{sec:sed-modeling-results} rather than from data.  The blue curve from Fig.~\ref{fig:deproj} is again shown for reference.\label{fig:sed-deproj}}
  \end{figure*}

  To verify how well the two components needed to fit the SED match the observational result, we used the output of the model shown in Fig.~\ref{fig:inner_disk} and applied the same procedure as used to derive Fig.~\ref{fig:deproj}. Because of the lack of a high-fidelity PSF model, no central star was added to the model data and the absolute calibration was performed such that the integrated flux of the B1 ring matches that measured on the data. The B1 component is matched almost perfectly, while B3 appears still visible, although the profile plot shows that in the model it appears less pronounced and slightly further in. All in all, it demonstrates a reasonable agreement, given that we did not actually inject the model into real data and run it through the analysis pipeline.

  B2 is unseen in the model, but as already mentioned in Sec.~\ref{sec:sed-modeling-results} it is not very bright anyway and degeneracies between dust properties and the dust distribution prohibits the attempt to fit an overly complex model that could reproduce the exact intensity distribution.

\end{appendix}

\end{document}